\documentclass[fleqn,usenatbib]{mnras}

\usepackage{newtxtext,newtxmath}

\usepackage[T1]{fontenc}

\DeclareRobustCommand{\VAN}[3]{#2}
\let\VANthebibliography\thebibliography
\def\thebibliography{\DeclareRobustCommand{\VAN}[3]{##3}\VANthebibliography}

\usepackage{graphicx}	
\usepackage{amsmath}	
\usepackage{comment}

\newcommand{\kms}{\mbox{km~s$^{-1}$}}

\newcommand{\mh}{\mbox{$\text{H}_{2}$}}
\newcommand{\brg}{\mbox{Br-$\gamma$}}

\title[The evaporating HH~900 outflow]{Illuminating evaporating protostellar outflows: ERIS/SPIFFIER reveals the dissociation and ionization of HH~900\thanks{Based on observations collected at the European Southern Observatory under ESO programmes 110.257T.001 and 0101.C-0391(A).}}

\author[M. Reiter et al.]{Megan Reiter,$^{1}$\thanks{E-mail: megan.reiter@rice.edu (MR)}
Thomas J. Haworth,$^{2}$
Carlo F. Manara,$^{3}$
Suzanne Ramsay,$^{3}$
Pamela D. Klaassen,$^{4}$ 
\newauthor
Dominika Itrich,$^{3}$
and Anna F. McLeod$^{5,6}$
\\
$^{1}$Department of Physics and Astronomy, Rice University, 6100 Main St - MS 108, Houston, TX 77005, USA \\
$^{2}$Astronomy Unit, School of Physics and Astronomy, Queen Mary University of London, London E1 4NS, UK \\ 
$^{3}$European Southern Observatory, Karl-Schwarzschild-Strasse 2, 85748 Garching bei M\"unchen, Germany \\
$^{4}$UK Astronomy Technology Centre, Royal Observatory Edinburgh, Blackford Hill, Edinburgh, EH9 3HJ, UK \\ 
$^{5}$Centre for Extragalactic Astronomy, Department of Physics, Durham University, South Road,  Durham DH1 3LE, UK\\
$^{6}$Institute for Computational Cosmology, Department of Physics, University of Durham, South Road, Durham DH1 3LE, UK
}

\date{Accepted 2023 October 24. Received 2023 October 13; in original form 2023 July 14}

\pubyear{2023}

\begin{document}
\label{firstpage}
\pagerange{\pageref{firstpage}--\pageref{lastpage}}
\maketitle

\begin{abstract}
Protostellar jets and outflows are signposts of active star formation. 
In H~{\sc ii} regions, molecular tracers like CO only reveal embedded portions of the outflow. 
Outside the natal cloud, outflows are dissociated, ionized, and eventually completely ablated, leaving behind only the high-density jet core. 
Before this process is complete, there should be a phase where the outflow is partially molecular and partially ionized. 
In this paper, we capture the HH~900 outflow while this process is in action. 
New observations from the ERIS/SPIFFIER near-IR integral field unit (IFU) spectrograph 
using the K-middle filter ($\lambda$=2.06–2.34~\micron) 
reveal \mh\ emission from the dissociating outflow and \brg\ tracing its ionized skin. 
Both lines trace the wide-angle outflow morphology but \mh\ only extends $\sim$5000~au into the H~{\sc ii} region while \brg\ extends the full length of the outflow ($\sim$12,650~au), indicating rapid dissociation of the molecules. 
\mh\ has higher velocities further from the driving source, consistent with a jet-driven outflow.  
Diagnostic line ratios indicate that photoexcitation, not just shocks, contributes to the excitation in the outflow. 
We argue that HH~900 is the first clear example of an evaporating molecular outflow and predict that a large column of neutral material that may be detectable with ALMA accompanies the dissociating molecules. 
Results from this study will help guide the interpretation of near-IR images of externally irradiated jets and outflows such as those obtained with the \emph{James Webb Space Telescope (JWST)} in high-mass star-forming regions where these conditions may be common. 

\end{abstract}




\section{Introduction}

Protostellar jets and outflows are produced by actively accreting young stellar objects (YSOs).  
Fast, collimated jets launched close to the YSO are often seen at visual wavelengths in hydrogen recombination and forbidden emission lines that are excited in shocks. 
Molecular emission lines in the millimetre (mm) and infrared (IR) trace outflows that may be launched directly from the circumstellar disk or entrained from the surrounding medium by an underlying jet. 
Jets and outflows may coexist and which component is more prominent / visible depends on the environment and evolutionary stage of the source \citep[see][for a recent review]{bally2016}.

A few well-studied examples like HH~46/47 have played an important role in understanding the relationship between jets and outflows. 
HH~46/47 is a parsec-scale bipolar outflow \citep{stanke1999}. 
The fast, collimated jet that emerges from an embedded young stellar object (YSO) is well-studied in bright recombination and forbidden lines that trace shock-excited gas in the jet body \citep[e.g.,][]{hea96,hartigan2005,hartigan2011,erkal2021}. 
On the opposite side of the embedded driving source, a classic molecular outflow propagates into the natal cloud with a wide-angle morphology traced by both the \emph{Spitzer Space Telescope} \citep{nor04} and the Atacama Large Millimeter Array \citep[ALMA;][]{arc13,zhang2016}.

Most of our knowledge about jets and outflows comes from sources like HH~46/47 -- those located in nearby star-forming regions that are forming primarily or exclusively low-mass stars. 
In regions with high-mass stars, the observational picture may look quite different. 
Copious UV photons from the most massive stars illuminate the surrounding cloud, including jets and outflows that emerge into the H~{\sc ii} region.  
External irradiation renders the entire jet body visible \citep{bal06,smith2010}, unlike jets in more quiescent regions where the only material emitting at visual and infrared wavelengths has been heated and excited in shock fronts \citep[e.g.,][]{bal01}. 
This is a distinct advantage for measuring the mass-loss rate (and thus accretion history of the driving source) because it can be determined using the physics of photoionized gas rather than complex, non-linear, and time-dependent shock models \citep{rei98,bally2006}.

Once jets and outflows emerge into the H~{\sc ii} region, molecules will quickly be dissociated. 
Traditional molecular outflow tracers, like CO, will only be seen in portions of the flow that remain embedded in the cloud \citep[e.g.,][]{cortes-rangel2020}. 
Outside the cloud, the outflow will be dissociated, ionized, and finally, completely ablated, leaving behind only the high-density jet core. 
Before this process is complete, there should be a phase where the molecular outflow is only partially dissociated with an ionized skin. 
The idea of an ionized outflow has been proposed for a few sources in the Carina Nebula \citep[HH~666 and HH~900;][]{smith2004_hh666,hartigan2015,reiter2015_hh666,reiter2015_hh900}. 
So far, there have been no unambiguous detections of a molecular outflow that extends into the H~{\sc ii} region.

\begin{figure}
    \centering
    \includegraphics[width=\columnwidth]{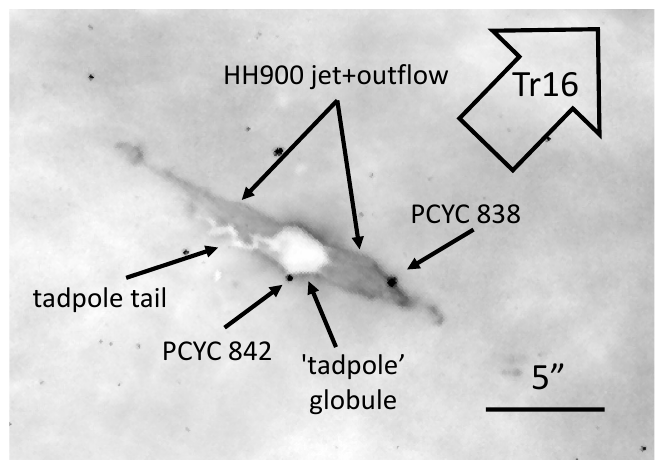}
    \caption{An H$\alpha$ image from \emph{HST} showing the HH~900 jet+outflow. 
    HH~900 emerges from a tadpole-shaped globule that is seen in silhouette against the bright background of the H~{\sc ii} region. The jet-driving source is unseen inside the opaque globule. Two additional candidate YSOs are seen outside the globule, PCYC~838 and PCYC~842. }
    \label{fig:context}
\end{figure}

In this paper, we present the first clear case of an evaporating molecular outflow. 
The HH~900 jet+outflow, emerges from a small, tadpole-shaped globule located in the heart of the Carina Nebula (see Figure~\ref{fig:context}). 
Dozens of O- and B-type stars in the nearby Trumpler~16 (Tr16) star cluster illuminate the system and may have triggered its formation \citep{reiter2020_tadpole_comp}. 
H$\alpha$ emission traces a wide-angle bipolar outflow that emerges from the opaque globule \citep{smith2010}. 
Unusually for an outflow, the H$\alpha$ emission appears to taper as it gets further from the driving source. 
This is the morphology expected for a jet-driven outflow \citep[see, e.g.,][]{arc07}, consistent with H$\alpha$ tracing the ionized skin of the outflow driven by an underlying jet (seen in [Fe~{\sc ii}]; \citealt{reiter2015_hh900,reiter2019_tadpole}).

CO observations with ALMA reveal a bipolar molecular outflow that extends only to the edge of the globule where it abruptly ends \citep{reiter2020_tadpole}.
In seeing-limited near-IR narrowband images, H$_2$ emission extends from the globule edge along the outflow axis \citep{hartigan2015}. 
Optical integral field unit spectroscopy from the Multi-Unit Spectroscopic Explorer (MUSE) revealed extended [C~{\sc i}] emission from the same region as the \mh\ \citep{reiter2019_tadpole}. 
[C~{\sc i}] is often observed to be coincident with H$_2$ in partially molecular gas \citep[e.g.,][]{escalante1991}, suggesting that the region where both lines are detected traces the transition between the fully molecular and fully ionized portions of the outflow.   
Together, these observations strongly suggest that HH~900 has a dissociating molecular outflow.

To test this hypothesis, we obtained new near-IR integral field unit spectroscopic observations from the Enhanced Resolution Imager and Spectrograph (ERIS) on the Very Large Telescope (VLT). 
These high spatial and spectral resolution observations allow us to probe the three key features we expect if HH~900 has a dissociating molecular outflow, namely: 
(1) the same morphology in molecular (\mh) and ionized (\brg) gas; 
(2) similar kinematics in molecular (\mh) and ionized (\brg) gas that are distinct from the fast, collimated jet \citep{reiter2015_hh900}; and 
(3) gas that is primarily photoexcited. 
These data provide comparable angular resolution and spatial coverage to previous observations obtained with the \emph{Hubble Space Telescope (HST)}/ACS, VLT/MUSE, and ALMA, providing a comprehensive view of the outflow.

\begin{figure}
    \centering
    \includegraphics[width=\columnwidth]{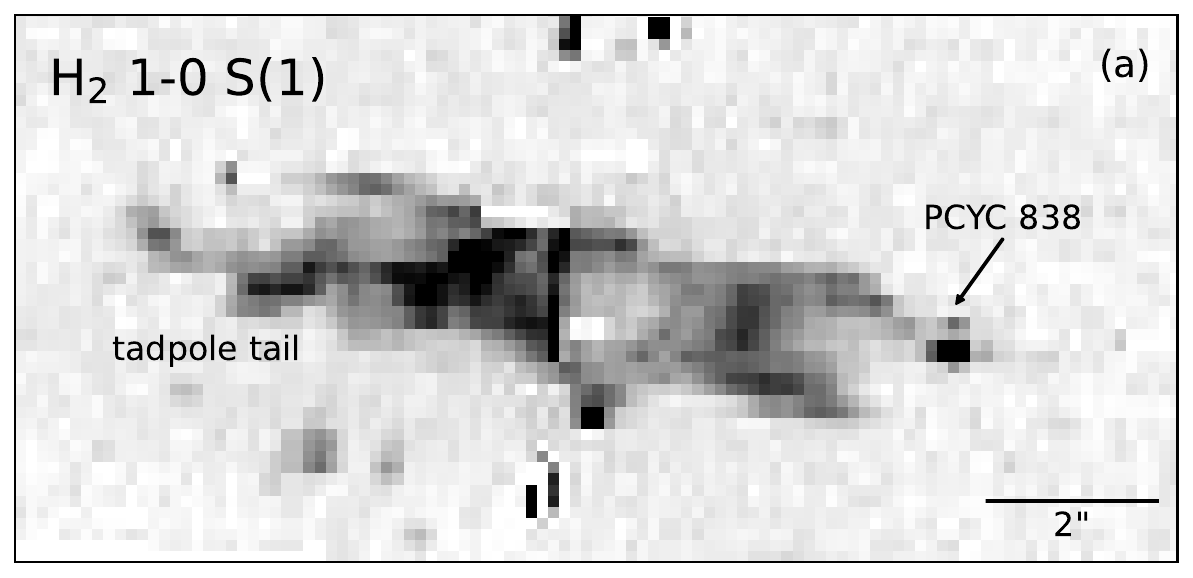}
    \includegraphics[width=\columnwidth]{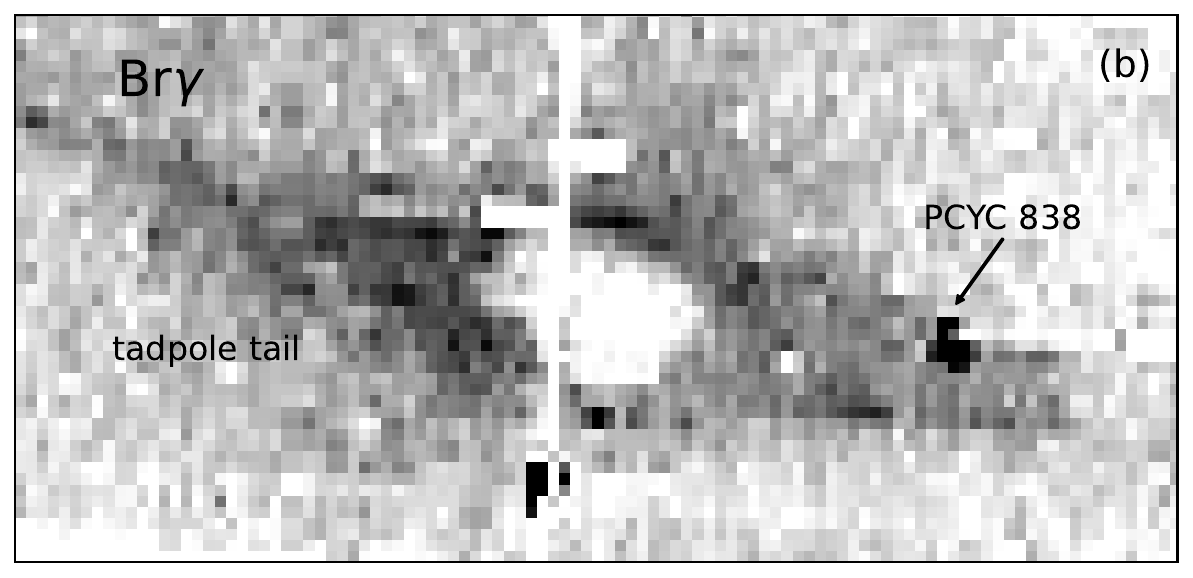}
    \includegraphics[width=\columnwidth]{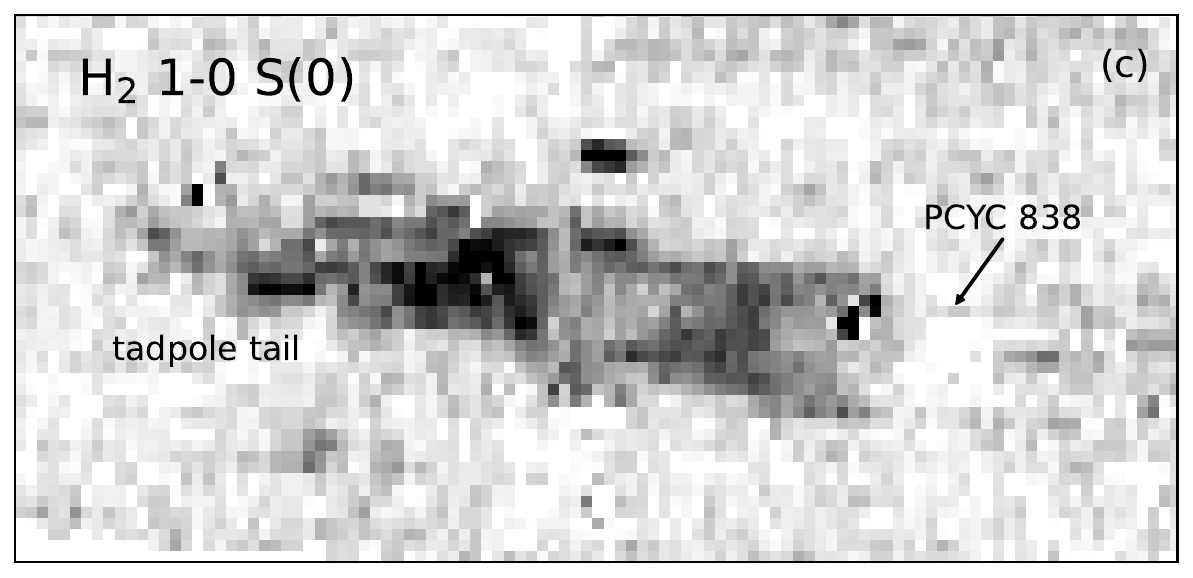}
    \includegraphics[width=\columnwidth]{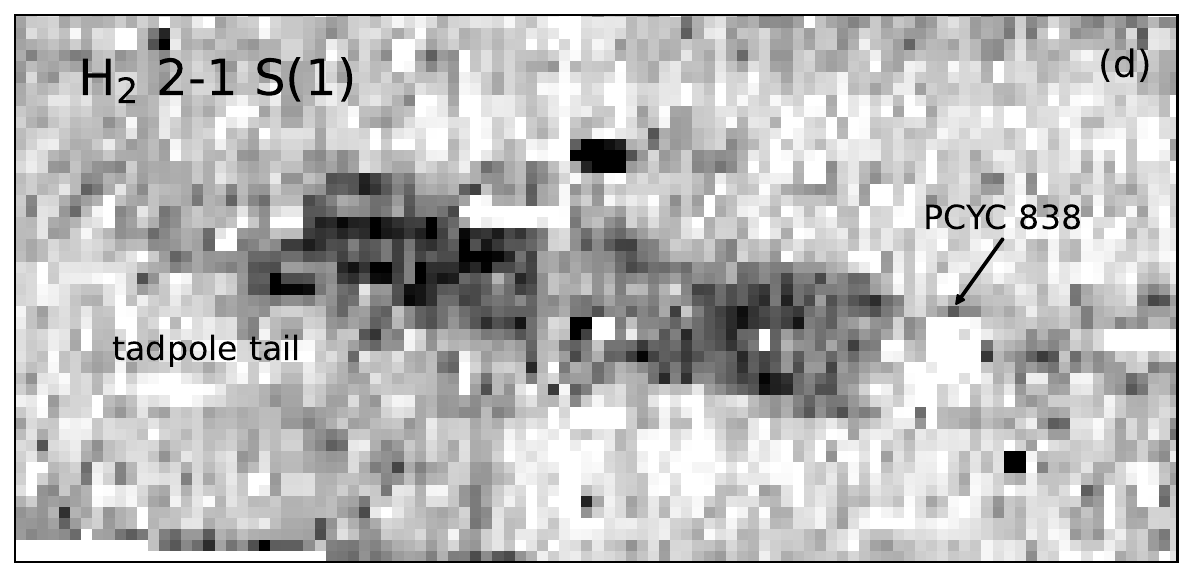}
    \caption{Integrated intensity (moment~0) maps of the lines detected in the ERIS SV data (see Table~\ref{t:lines}): 
    (a) \mh\ 1-0 S(1); 
    (b) \brg; 
    (c) \mh\ 1-0 S(0); 
    and 
    (d) \mh\ 2-1 S(1). } 
    \label{fig:moment0_maps}
\end{figure}

\section{Observations}\label{s:data}

\subsection{SPIFFIER/ERIS}
We observed HH~900 (RA=10:45:19.3, Dec=$-$59:44:23) using the recently-commissioned SPIFFIER near-IR integral field unit spectrograph of the VLT/ERIS instrument \citep{davies2018,davies2023}. 
SPIFFIER is the refurbished integral field unit spectrograph SPIFFI (SPectrometer for Infrared Faint Field Imaging) that was previously part of 
the Spectrograph for INtegral Field Observations in the Near Infrared (SINFONI).  
Data were obtained as part of the ERIS Science Verification
\footnote{\url{https://www.eso.org/sci/activities/vltsv/erissv.html}} for Pr. Id. 110.257T (PI: M.\ Reiter) on the night of  
05 December 2022.

We used the lowest resolution plate scale which provides $125 \times 250$~mas spatial pixels (spaxels) over an $8^{\prime\prime} \times 8^{\prime\prime}$ field-of-view. 
Two overlapping pointings capture the entire extent of the HH~900 jet+outflow (excluding distant bow shocks). 
Observations were obtained in good weather with seeing $\lesssim$0.8\arcsec. 
We also utilized laser guide star (LGS) adaptive optics (AO) correction to further improve the image quality.

We used the high-resolution K-middle filter ($\lambda$=2.06--2.34~$\mu$m) which covers the H$_2$ line at 2.12~$\mu$m and Br$\gamma$ at 2.16~$\mu$m simultaneously with spectral resolution R$\sim$11,200. 
The corresponding velocity resolution is $\sim$25~\kms.

Data were reduced using a beta version of the ERIS pipeline provided by ESO at the time of the Science Verification run and executed through the ESO Reflex workflow \citep{reflex}. The pipeline corrects the data from the instrumental signatures, i.e. darks and flats, calibrates a wavelength solution using the associated arc lamps, applies a field distortion mapping, computes and substracts the sky contribution, and generates a resampled 3D data cube. This is then used in the following analysis.

The systemic velocity of HH~900 was measured by \citet{reiter2020_tadpole} who found $v_{\mathrm{LSR}} = -33.5$~\kms. 
For the coordinates of Carina, 
$v_{\mathrm{helio}} \approx v_{\mathrm{LSR}} + 11.6$~\kms\ \citep{kiminki2018}. 
Using this, we compute a heliocentric velocity of $v_{\mathrm{helio}} = -21.9$~\kms\ for HH~900. 
All velocities in this paper are reported relative to the heliocentric velocity of HH~900. 

\subsection{Archival data} 

To align the ERIS/SPIFFIER data with the archival images, we use header astrometry for preliminary registration, then apply additional linear offsets to align the point sources near the globule and outflow. 
The spatially-resolved globule and tadpole tail provide a second check for data with few point sources (i.e.\ from ALMA). 
Data were additionally rotated and shifted to minimize subtraction residuals. 
Typical alignment uncertainties are on the order of a pixel ($\sim$0.15\arcsec).

\subsubsection{ \emph{HST} H$\alpha$ and [Fe~{\sc ii}]}
A narrowband H$\alpha$ image was obtained with the F658N filter on the Advanced Camera for Surveys (ACS) on 04 August 2014 (programme GO-13390, PI: N.\ Smith). 
The narrowband [Fe~{\sc ii}] 1.64~\micron\ (F164N) and offline continuum (F167N) images were obtained with the infrared channel of the Wide-Field Camera 3 (WFC3-IR) on 28 December 2013 (programme GO-13391, PI: N.\ Smith). 
Details of these data and their reduction are presented in \citet{reiter2015_hh900}.

\subsubsection{MUSE}
The HH~900 system was observed with the Multi-Unit Spectroscopic Explorer (MUSE) integral field unit spectrograph on the VLT 03 April 2018 (programme ID 0101.C-0391(A); PI: M.\ Reiter). 
These observations utilised the GALACSI Adaptive Optics module in Wide Field Mode (WFM) to provide $\sim$0.8\arcsec\ angular
resolution over the 1\arcmin $\times$ 1\arcmin\ field-of-view. 
MUSE provides spectral coverage from 4650--9300\AA\ with a gap between $\sim$5800--5950\AA\ for the laser guide stars with spectral resolution $R=2000-4000$. 
Details of those observations may be found in \citet{reiter2019_tadpole}.

\subsubsection{ALMA}
ALMA Band~6 observations of the HH~900 system were obtained on 
08 May 2017 and 25 September 2017 using medium and long baseline configurations (programme ID 2016.1.01537.S, PI: A.\ Guzm\'{a}n). 
The maximum angular resolution is 0.02\arcsec, comparable to the resolution of H$\alpha$ images from \emph{HST}. 
CO lines were observed with a velocity resolution of 0.08~\kms. 
Anaylsis and details may be found in 
\citet{reiter2020_tadpole}.

\section{Results}\label{s:results} 

\begin{table}
\caption{Primary emission lines detected in the HH~900 jet+outflow. Wavelengths for the \mh\ lines are from \citet{levenson2000}; \brg\ wavelength is from \citet{chang1996observation} via NIST\protect\footnotemark{}. All wavelengths are in vacuum. 
\label{t:lines} 
}
\centering
\begin{tabular}{lc} 
		\hline
		Line & wavelength \\ 
         & [\micron] \\ 
		\hline
		\mh\ 1-0 S(1) & 2.12183 \\ 
		\brg\ & 2.16612 \\ 
		\mh\ 1-0 S(0) & 2.22329 \\ 
        \mh\ 2-1 S(1) & 2.24772 \\ 
		\hline
\end{tabular}
\end{table}
\footnotetext{ \href{ https://www.nist.gov/ }{ https://www.nist.gov/ }}

We detected \brg\ and 3 bright \mh\ lines with ERIS/SPIFFIER (see Table~\ref{t:lines} and Figure~\ref{fig:moment0_maps}). 
All of the detected lines are spatially extended, tracing the bipolar HH~900 outflow. 

Bright emission from all lines also trace the ionization front on the globule surface. 
The tadpole tail is prominent in \mh\ images, tracing the same morphology seen in silhouette in H$\alpha$ \citep{smith2010} and in emission in CO \citep{reiter2020_tadpole}.

\begin{figure}
    \centering
    \includegraphics[width=\columnwidth]{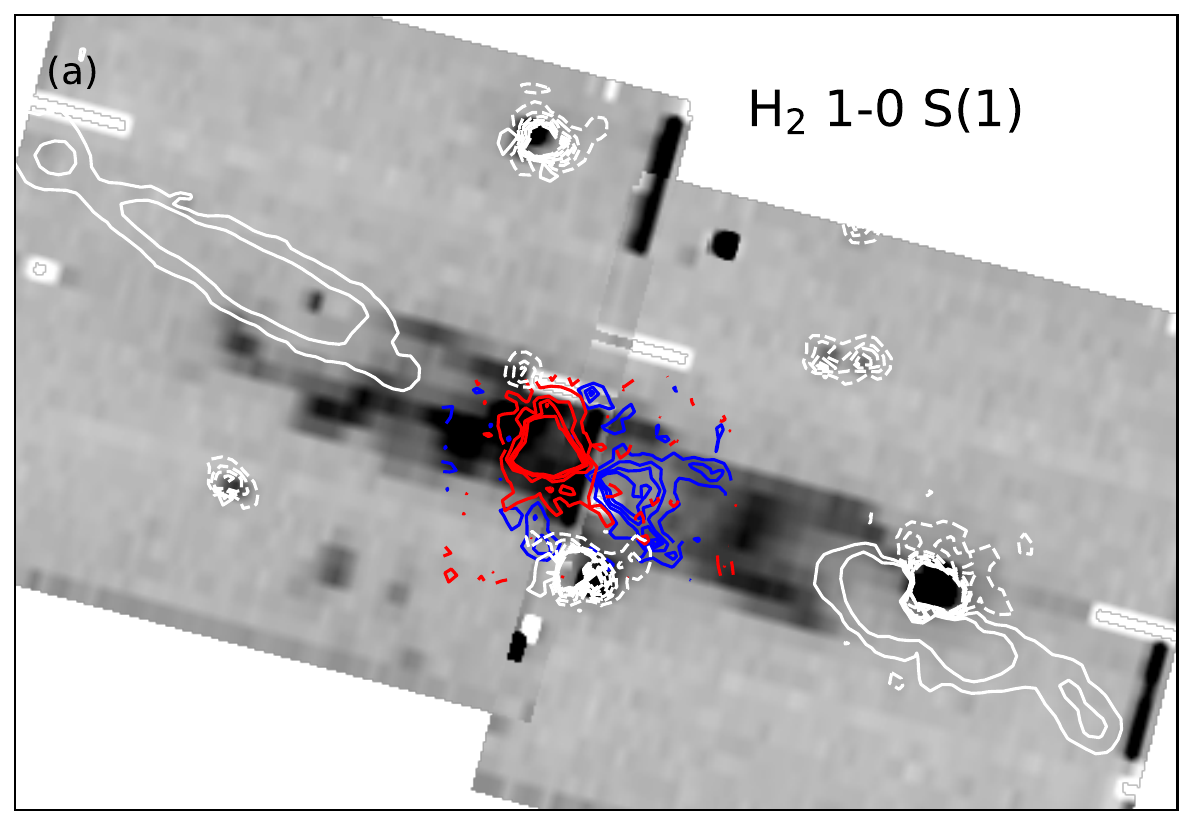}
    \includegraphics[width=\columnwidth]{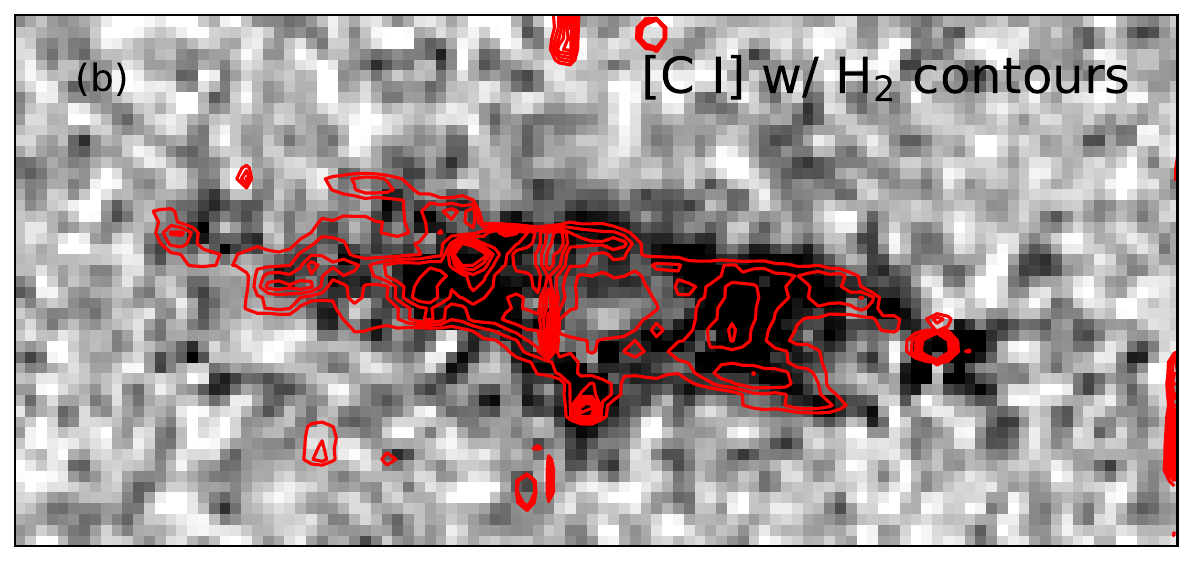}
    \caption{\textbf{Top:} Grayscale image of \mh\ from ERIS; 
    red and blue contours show CO J=2-1 from ALMA (emission is integrated from $-$30~\kms\ to $-$11~\kms\ and $-$57~\kms\ to $-$35.75~\kms, respectively); white contours show [Fe~{\sc ii}] 1.64~\micron\ from \emph{HST}. 
    \textbf{Bottom:} \mh\ contours (red; 10--50$\sigma$ in steps of $5\sigma$) on a [C~{\sc i}] 8727~\AA\ image from MUSE. 
    }
    \label{fig:muse_alma_comps}
\end{figure}

\begin{figure*}
    \centering
    \includegraphics[width=\textwidth]{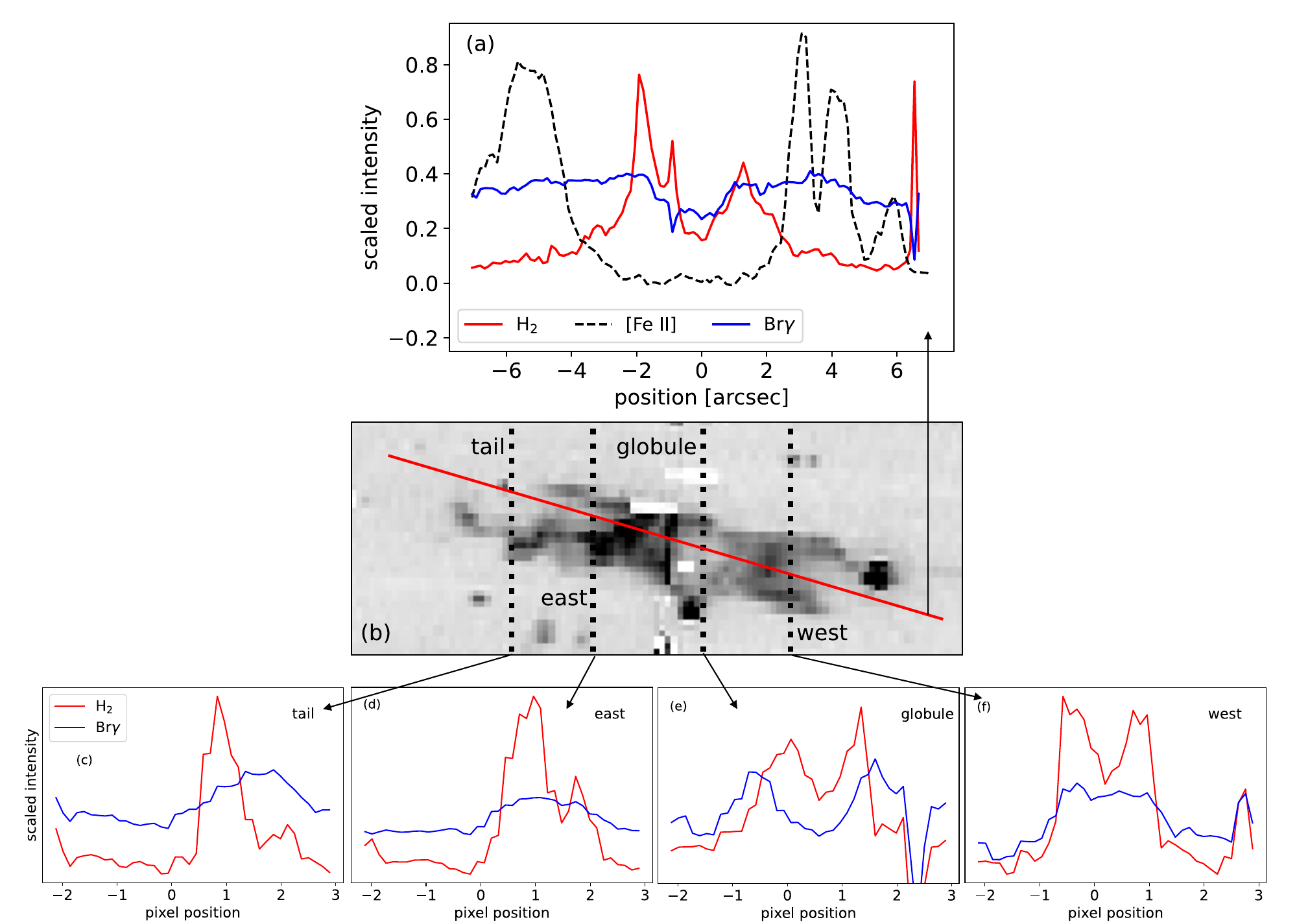}
    \caption{\textbf{Top:} Intensity tracings showing \mh\ (red), \brg\ (blue), and [Fe~{\sc ii}] (black dashed line) along the length of the outflow \textbf{(a)}. \mh\ extends a small distance from the globule but tapers off before [Fe~{\sc ii}] from the jet is first detected. \brg\ is bright throughout the length of the outflow. 
    \textbf{Middle:} \mh\ image with lines showing the location of the hortizontal (red line) and vertical (black dotted lines) intensity tracing locations \textbf{(b)}.
    \textbf{Bottom:} Vertical intensity tracings ordered from east to west.  
    In the eastern-most slice \textbf{(c)}, both lines trace the outflow but \mh\ is dominated by emission from the tadpole tail. 
    Closer to the globule \textbf{(d)}, \mh\ and \brg\ trace the wide-angle outflow. 
    A tracing through the globule itself \textbf{(e)} reveals \mh\ offset inside \brg, consistent with a steady-state photodissociation region on the surface of a post-collapse globule \citep[see][]{reiter2020_tadpole_comp}. 
    The western-most slice through the HH~900 outflow \textbf{(f)} shows the wide opening angle traced by both components and illustrates the limb-brightening of the \mh\ flow. 
     }
    \label{fig:intensity_tracing}
\end{figure*}

Both \brg\ and \mh\ emission from HH~900 trace the wide-angle outflow, extending smoothly from where the CO outflow ends at the edge of the globule \citep[see Figure~\ref{fig:muse_alma_comps} and][]{reiter2020_tadpole}. 
\brg\ emission traces the same morphology as H$\alpha$, as expected. 
In the H~{\sc ii} region, \mh\ traces a broadly similar morphology to the ionized outflow traced by H$\alpha$ and \brg\ but the surface brightness is less uniform. 
\mh\ appears limb-brightened with a measurable dip in the intensity at the mid-point of the outflow lobe (see Figure~\ref{fig:intensity_tracing}). 
\mh\ emission is less extended, reaching only $\sim 2.2$\arcsec\ (0.02~pc) from the globule, compared to $\sim 5.5$\arcsec\ (0.06~pc) for H$\alpha$ and \brg. 
This is the same extent seen in seeing-limited \mh\ images from \citet{hartigan2015} and coincides with the [C~{\sc i}] emission seen with MUSE \citep[see Figure~\ref{fig:muse_alma_comps} and][]{reiter2019_tadpole}. 

In the following analysis, we refer to the \mh\ 2.12~\micron\ line simply as `\mh' and specify the transition of the other \mh\ lines where used.

%
%
\subsection{Velocity structure}\label{ss:velocities}
%
\begin{figure}
    \centering
    \includegraphics[scale=0.605,trim=0mm 63mm 0mm 25mm]{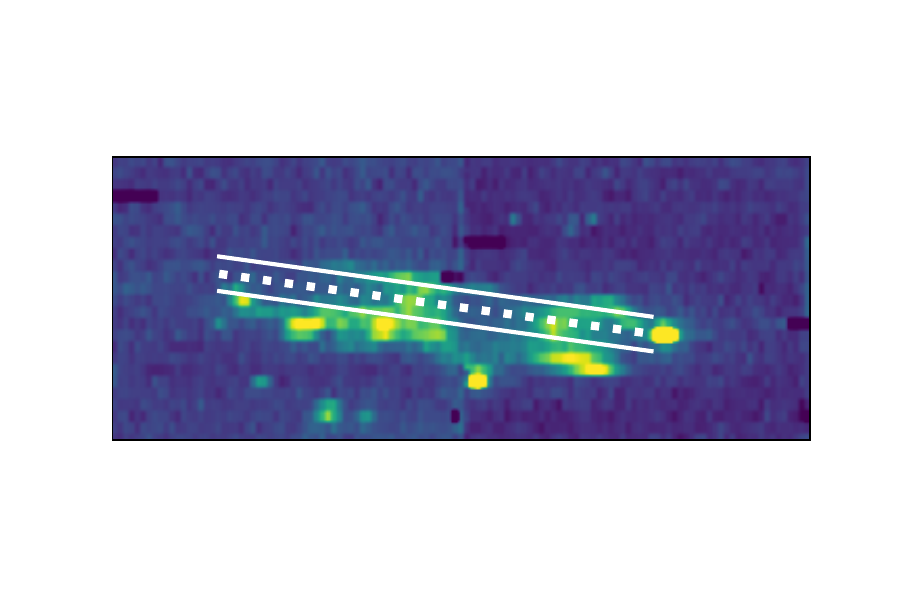}
    \includegraphics[scale=0.605,trim=0mm 75mm 0mm 0mm]{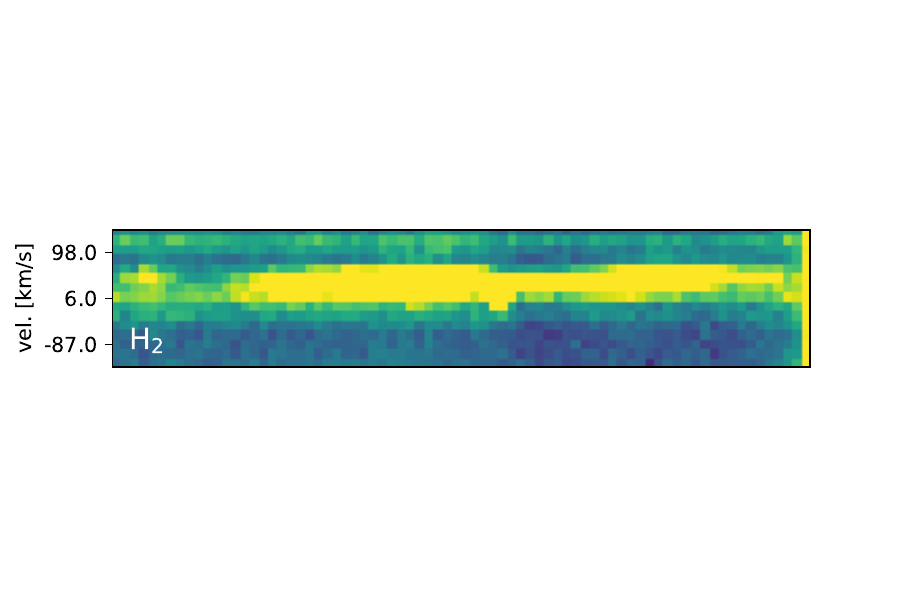}
    \includegraphics[scale=0.605,trim=0mm 75mm 0mm 0mm]{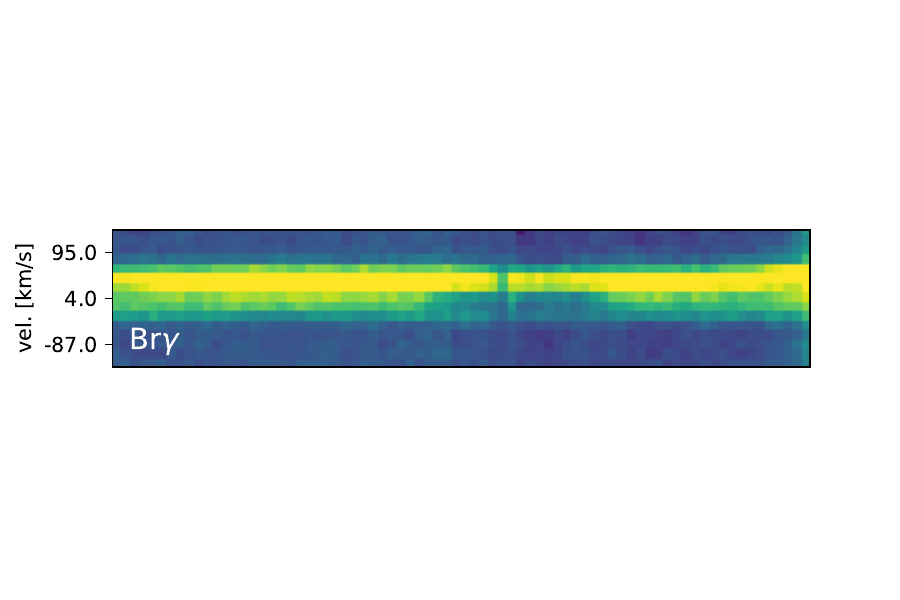}
    \includegraphics[scale=0.605,trim=0mm 25mm 0mm 0mm]{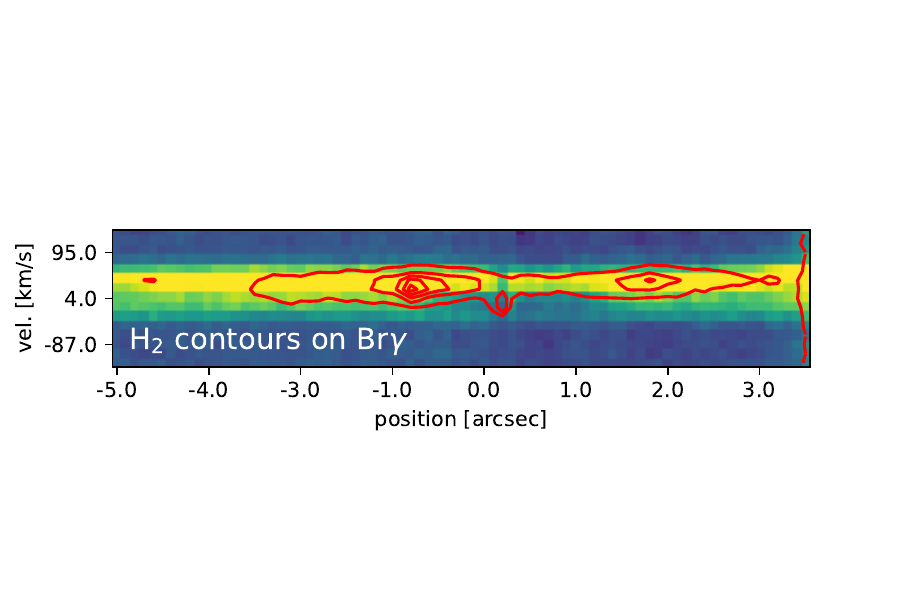}
    \includegraphics[scale=0.575]{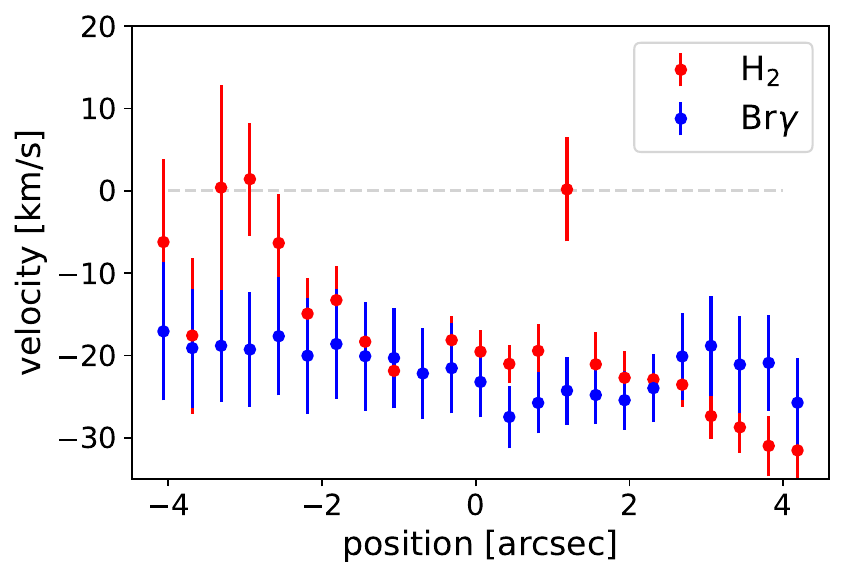}
    \caption{\textbf{Top:} \mh\ image showing the location and width of the slice used to make the P-V diagrams. 
    \textbf{Middle:} \mh\ and \brg\ P-V diagrams and a third P-V diagram showing \mh\ contours (red) on the \brg\ P-V diagram. 
    \textbf{Bottom:} The best-fit velocity derived from fitting a Gaussian to velocity slices along the jet axis (see Section~\ref{ss:velocities}). }
    \label{fig:pv}
\end{figure}
The HH~900 outflow lies close to the plane of the sky, with a tilt angle $\lesssim 10^{\circ}$ \citep{reiter2015_hh900}. 
The bulk of the outflow velocity is therefore captured in the tangential motions in the plane of the sky. 
We benefit from the high velocity resolution of SPIFFIER/ERIS to measure the more modest radial velocities from the outflow. 
To measure the velocity of the outflow components traced by \mh\ and \brg, we construct position-velocity (P-V) diagrams of each line (see Figure~\ref{fig:pv}). 
We extract a slice 5 pixels wide through the center of the jet+outflow. 
Both H$_2$ and \brg\ show blueshifted emission from the western limb of the jet and redshifted emission to the east, consistent with other velocity measurements of the jet+outflow \citep{reiter2015_hh900,reiter2020_tadpole}.

To obtain a more precise estimate, we fit a Gaussian to the velocity profile in different slices across the outflow.  
We take the average emission in three-pixel-wide slices (the equivalent of three columns in the P-V diagrams). 
Line profiles are largely single-peaked. 
Low \mh\ velocities between $-3$\arcsec\ and $-4$\arcsec\ (on the eastern side of the globule) are contaminated by the blueshifted tadpole tail (see Appendix~\ref{s:slices} and Figure~\ref{fig:3_slices}). 
Note that the tadpole tail is blueshifted on the same (eastern) side of the globule where the outflow is redshifted. 
Overall, \mh\ velocities in the outflow are faster than \brg.

Velocity measurements for the HH~900 jet+outflow were previous reported by \citet{reiter2015_hh900}. 
\brg\ velocities are consistent with the marginally-resolved blueshifted velocities seen in the western limb of the outflow in H$\alpha$ (which reach a maximum blueshifted velocity of $\sim$-16~\kms). 
All outflow velocities traced by \mh\ and \brg\ are slower than the [Fe~{\sc ii}] emission from the jet ($\pm \sim$30~\kms; see Appendix~\ref{s:vel_WTF}). 

\subsection{The YSOs}

The HH~900 driving source remains unseen at near-IR wavelengths as it is deeply embedded inside the tadpole-shaped globule. 
However, we detect two point sources that are visible outside the globule (see Figure~\ref{fig:context}). 
Both were identified as candidate young stellar objects (YSOs) in the Pan Carina YSO Catalog (PCYC) by \citet{povich2011}. 
We use aperture extraction to isolate the stellar spectrum and remove the sky background. 
For both objects, the `sky' background includes nebular emission from the H~{\sc ii} region and structured emission from either the edge of the globule or the limb-brightened HH~900 outflow. 
As a result, the \brg\ and \mh\ line profiles may be contaminated with residual nebular emission. 
Spectra of both sources are shown in Appendix~\ref{s:pcyc842}. 

The first source, PCYC~838, lies directly on top of the western limb of the HH~900 jet+outflow. 
We detect \brg\ in emission in the PCYC~838~YSO spectrum but no \mh\ lines. 
\brg\ emission may indicate active accretion \citep[e.g.,][]{fairlamb2017}. 
However, we do not detect other indicators of the circumstellar disk such as the CO bandhead in emission or absorption \citep[e.g.,][]{carr1989,calvet1991}. 
The absence of the molecular bands may point to a slightly higher mass for this object, consistent with the spectral type (roughly G-type) estimated from the optical spectrum \citep{reiter2019_tadpole} and the mass estimated from the best-fitting model of the spectral energy distribution \citep[$2.5 \pm 1.2$~M$_{\odot}$;][]{povich2011}.

A second star, identified as candidate YSO PCYC~842 by \citet{povich2011}, lies at the bottom of the tadpole-shaped globule. 
No prominent emission lines (i.e.\ \brg) are detected in this source. 
No millimeter continuum or molecular line emission associated with this source was detected with ALMA \citep{reiter2020_tadpole}, suggesting a lack of circumstellar material. 
Indeed, \citet{reiter2015_hh900} questioned whether this source is a YSO based on its motion \textit{toward} the globule and the low resolution of the data used for its original classification. 
The relatively featureless spectrum of this source does not provide evidence that this source is young or actively accreting.

\section{HH~900 as an evaporating molecular outflow}

We argue that the morphology, velocity, and excitation of the extended \mh\ and \brg\ emission in HH~900 trace an ionized and evaporating outflow. 
In this picture, the molecular outflow is rapidly but not instantaneously dissociated once it emerges into the H~{\sc ii} region. 
Extended \mh\ emission traces the extent (and therefore the time) that molecules survive in the H~{\sc ii} region. 
Spatially coincident [C~{\sc i}] emission seen with MUSE supports this interpretation (see Figure~\ref{fig:muse_alma_comps}).
The ionization potential of carbon is lower than hydrogen (11.26~eV compared to 13.6~eV), so 
[C~{\sc i}] is expected to coexist with partially dissociated \mh\ \citep[see, e.g.,][]{osterbrock2006}.

High spatial and spectral resolution integral field unit spectroscopy with ERIS allows us to perform three key tests of this hypothesis. 
If HH~900 is an evaporating molecular outflow we expect: 
\textbf{(1)} the same morphology in molecules (\mh) and recombination lines (\brg);  
\textbf{(2)} similar outflow-like kinematics in molecular (\mh) and recombination lines (\brg) that are distinct from the fast, collimated jet traced by [Fe~{\sc ii}]; 
and  
\textbf{(3)} photoexcitation dominating over shock excitation in the outflow. 
We discuss each test individually below.

\subsection{Morphology}\label{ss:morphology}
\begin{figure}
	\includegraphics[width=\columnwidth]{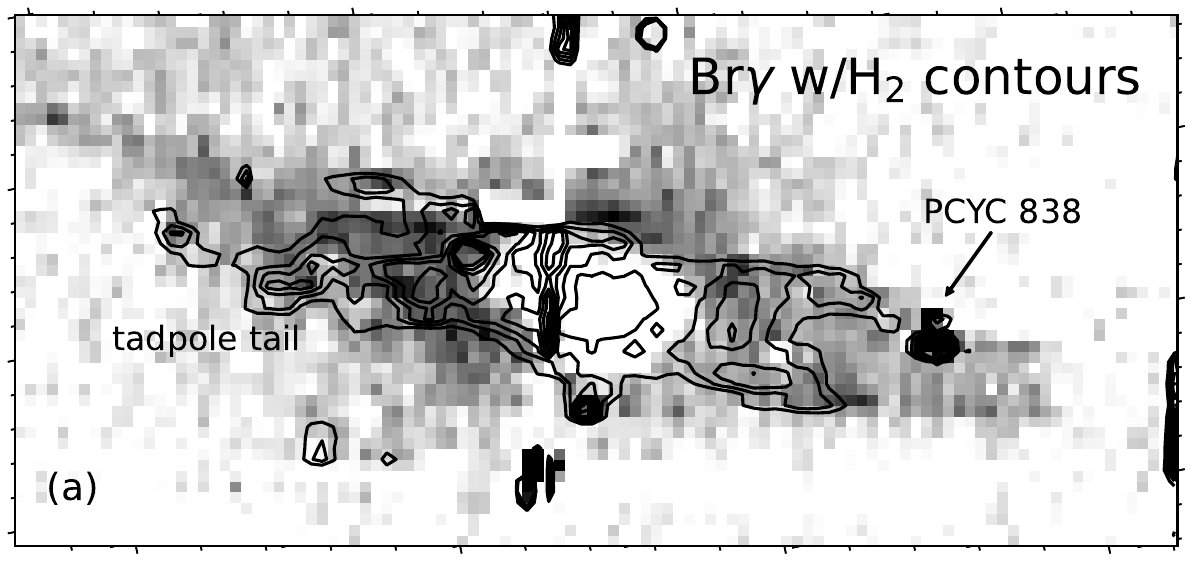}
    \includegraphics[width=\columnwidth]{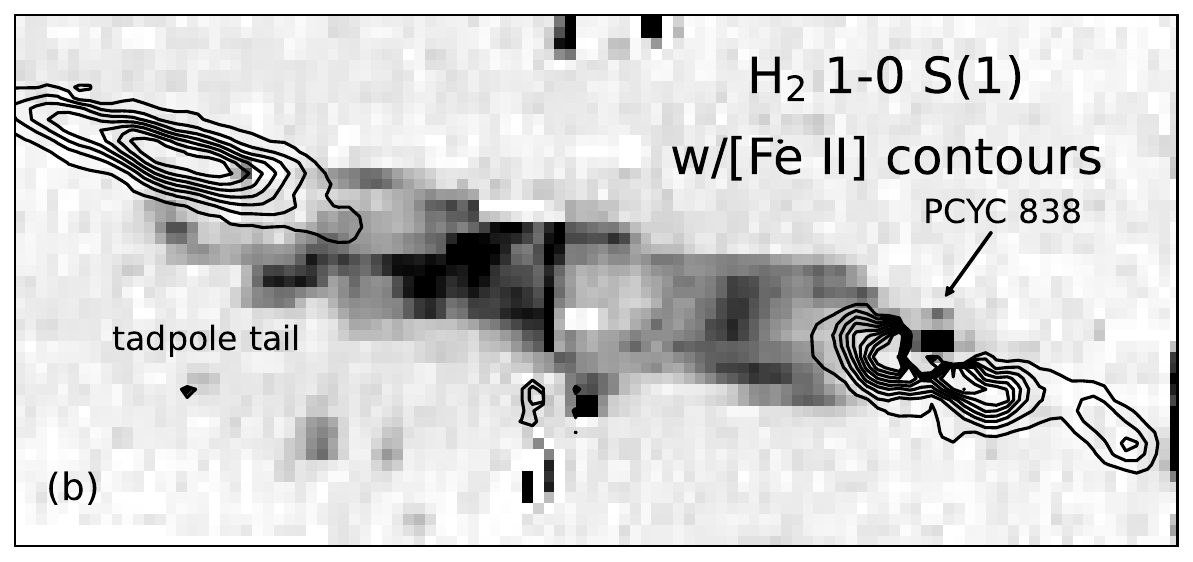}
 	\includegraphics[width=\columnwidth]{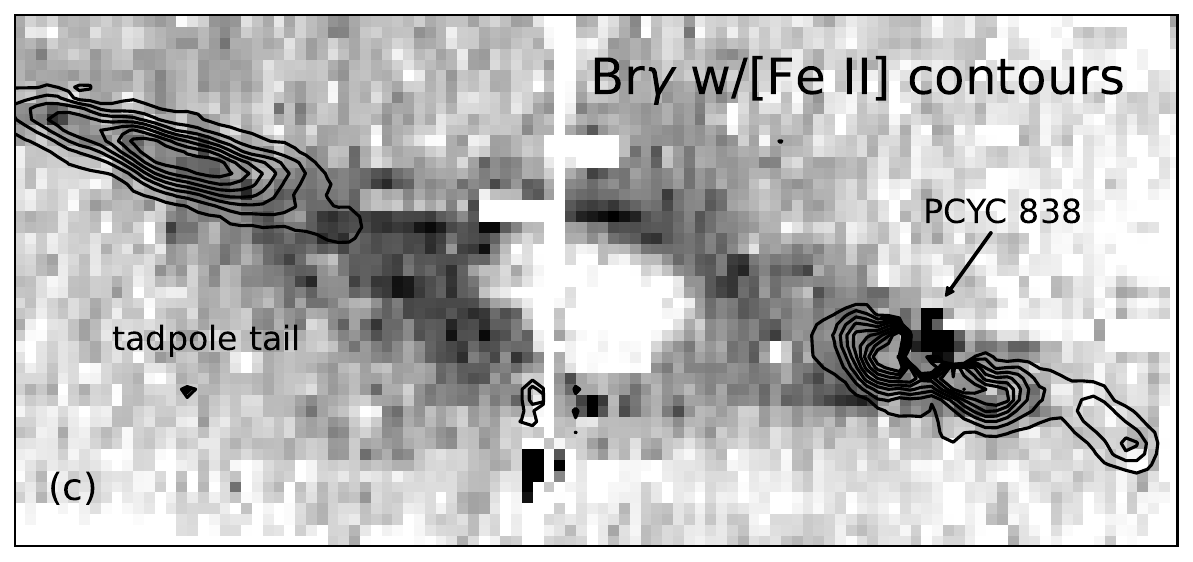}
    \includegraphics[width=\columnwidth]{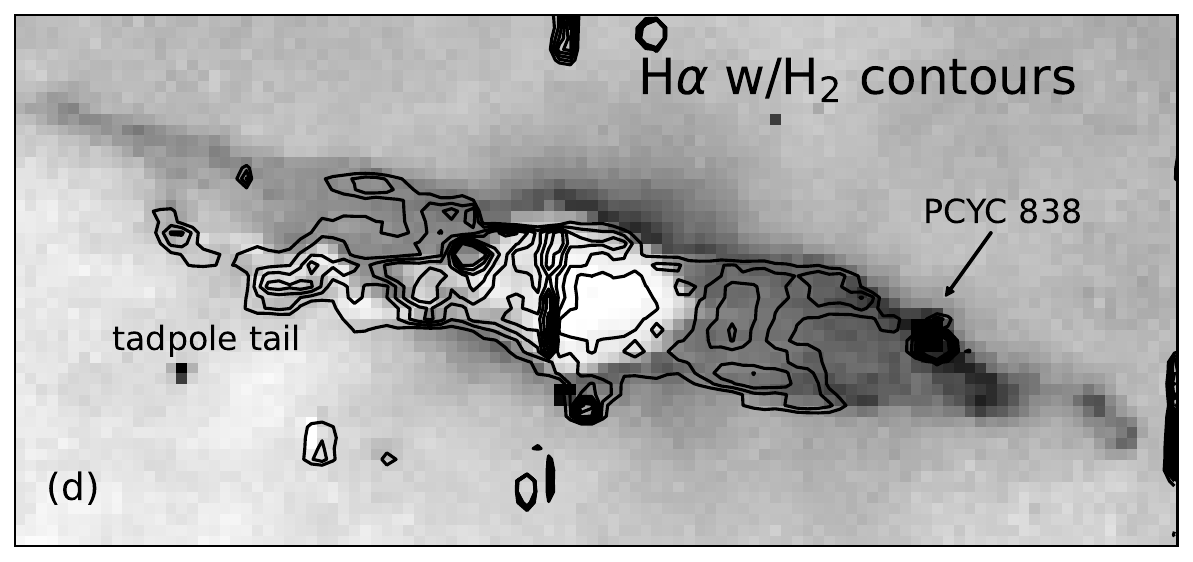}
    \caption{ \mh\ contours (black) on a \brg\ grayscale image \textbf{(a)}. 
    [Fe~{\sc ii}] 1.64~\micron\ contours on an \mh\ image \textbf{(b)} and \brg\ image \textbf{(c)}. 
    \mh\ contours on an H$\alpha$ image from \emph{HST} \textbf{(d)}. 
    \mh\ contours are 10--50$\sigma$ in steps of 5$\sigma$; [Fe~{\sc ii}] contours are 3--30$\sigma$ in steps of 3$\sigma$. 
    }
    \label{fig:hst_comp}
\end{figure}
\mh\ emission extends beyond the edge of the globule, tracing the same wide-angle flow as \brg\ (see Figure~\ref{fig:hst_comp}). 
The \mh\ appears limb-brightened, as though tracing the edges of an outflow cavity. 
The width of the \mh\ and Br$\gamma$ profiles are comparable for intensity tracings through the outflow (see Figure~\ref{fig:intensity_tracing}). 
This is in contrast to tracings through the globule itself that show an offset of $\sim$0.5\arcsec\ between the \brg\ and \mh\ emission peaks.

The key morphological difference between \mh\ and \brg\ is their extent. 
\mh\ reaches $\sim$2.2\arcsec\ (0.02~pc) from the globule edge into the H~{\sc ii} region.
The terminal edge of the limb-brightened \mh\ outflow coincides with the onset of [Fe~{\sc ii}] emission from the jet (see Figure~\ref{fig:hst_comp}). 
\brg\ extends $\sim$5.5\arcsec\ (0.06~pc) tracing the full length of the jet seen in H$\alpha$ and [Fe~{\sc ii}]. 
This is consistent with an evaporating molecular outflow if \mh\ disappears at the point where molecules are completely dissociated.

\subsection{Outflow velocities}\label{ss:vel_comp}
Outflow velocities traced by \mh\ increase with distance from the driving source. 
This Hubble-like flow is characteristic of jet-driven outflows \citep[see Figure~2 from][]{arc07}. 
The highest \mh\ velocities are $\pm \sim 20$~\kms, remarkably similar to the highest velocities seen from the embedded CO outflow \citep{reiter2020_tadpole}. 
\mh\ velocities are $\sim$10~\kms\ faster than \brg\ in both limbs of the outflow (see Figure~\ref{fig:pv}) and for all slices across the globule (see Appendix~\ref{s:slices}). 
This velocity difference is comparable to the sound speed in ionized gas ($c_s \sim 11$~\kms), the expected velocity of a photoevaporative flow. 

The fastest velocities measured in \mh\ and \brg\ are both lower than in the collimated [Fe~{\sc ii}] jet where  
\citet{reiter2015_hh900,reiter2019_tadpole} found fast, steady velocities up to $\pm$30~\kms. 
\brg\ outflow velocities are consistent with zero while the fastest \mh\ velocities are closer to 30~\kms. 
Several other outflows have been observed to have a nested or `onion-layer' structure with fast, highly collimated components on the jet axis surrounded by slower, wider angle layers \citep[e.g.][]{lavalley-fouquet2000,bac00,pyo03,coffey2008}. 
This same structure is seen in HH~900 with [Fe~{\sc ii}] emission tracing the fast core of the jet while \mh\ traces molecular layers either lifted from the disk or entrained from the envelope/globule. 
On the surface is the ionized \brg\ emission where the strong external radiation field first encounters the slowest, widest layers of the outflow.

\subsection{Excitation} 
\begin{figure}
    \centering
    \includegraphics[width=\columnwidth]{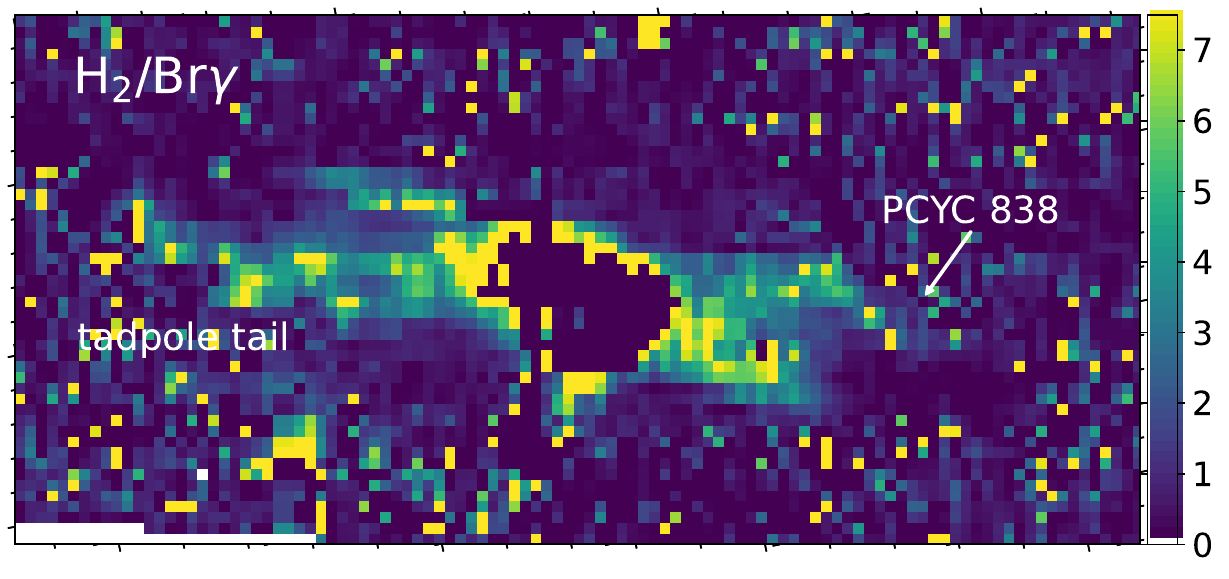}
    \includegraphics[width=\columnwidth]{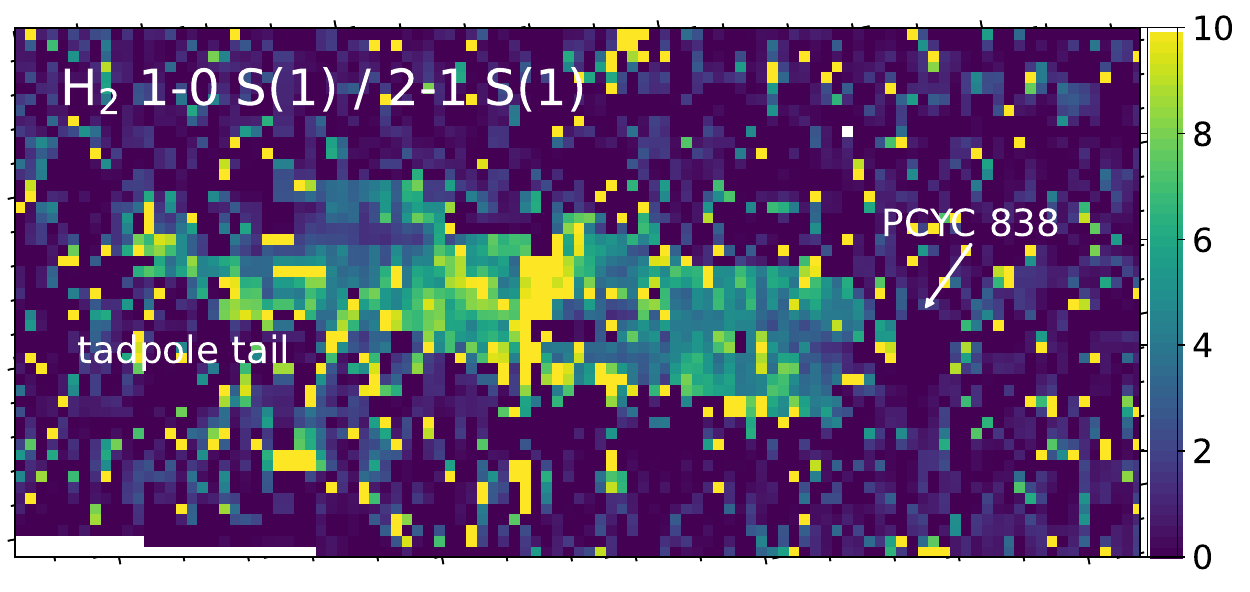}
    \caption{\textbf{Top:} Intensity map showing the \mh\ 1-0 S(1) / \brg\ line ratio. Values $<$1 indicate that photoexcitation dominates. 
     \textbf{Bottom:} Intensity maps of the \mh\ 1-0 S(1) / \mh\ 2-1 S(1) line ratio. The expected ratio is $\sim$3 in photoexcited gas, $\sim$10 in shock-excited gas.}
    \label{fig:line_ratios}
\end{figure}
We measure two different emission line ratios in the spatially-resolved outflow to determine the excitation of the gas. 
All four emission lines detected with these ERIS observations are within a narrow wavelength range so differences in wavelength-dependent extinction are negligible.  
We expect \mh/\brg$<$1 for a dissociating outflow where photoexcitation will play a dominant role \citep[the expected ratio in shock-excited gas is $>$1; see discussion in e.g.,][]{yeh2015}. 
The \mh\ 1-0 S(1) / \mh\ 2-1 S(1) ratio provides an additional diagnostic. 
For shock-excited gas, models predict a line ratio $\gtrsim 10$ \citep{shull1978}; for photo-excited gas, the expected line ratio is $\sim 1.5-3$ \citep{black1987}. 

Line ratio maps are shown in Figure~\ref{fig:line_ratios}. 
We take the ratio of the integrated intensity (moment 0) obtained by integrating each line over an interval $\pm\sim$4~\AA\ from the line center. 
Ambient nebular emission is estimated from the median value of the sky above and below HH~900.

The \mh/\brg\ ratio is $\geq 1$ throughout the HH~900 system. 
The highest values are seen on the eastern edge of the globule and in a few knots in the tadpole tail indicating that shock-excitation contributes in these portions of the system. 
Ratios in the outflow are lower but increase to higher values toward the outflow edges. 
No part of the system 
has \mh/\brg\ $< 1$. 

The interpretation of the \mh\ 1-0 S(1) / \mh\ 2-1 S(1) ratio map is less clear.
All values in the \mh\ 1-0 S(1) / \mh\ 2-1 S(1) ratio map are $<$10, below the range expected for shock-excited gas ($\gtrsim$10). 
However, throughout the map, the flux ratio is $>$3, larger than expected for pure photoexcitation. 
Unlike the \mh/\brg\ ratio map, the globule edge is not prominent in the \mh\ 1-0 S(1) / \mh\ 2-1 S(1) ratio map and there are no prominent bright spots in the tadpole tail.

Based on these ratio maps, the simplest interpretation is that a combination of shocks and photoexcitation contribute to the observed emission.  
Note that the typical \mh\ knot sizes seen in the spectacular HH~212 jet are $\sim$250$-$500~au \citep{zinnecker1998}. 
At the distance of Carina \citep[2.35~kpc;][]{goeppl2022}, this corresponds to $<0.25^{\prime\prime}$ -- too small to be resolved with ERIS.

\section{Discussion}\label{s:discussion}

Evaporating molecular outflows are the missing piece that connects (cold) molecular outflows seen inside clouds with the hot ionized jets laid bare in H~{\sc ii} regions. 
Once outflows leave the protection of their natal cocoons, external irradiation rapidly heats, dissociates, and ionizes them. 
This explains the abrupt end of CO outflows like HH~900 at the edge of the globule \citep[or dust pillar in the case of HH~901 and HH~902, see][]{cortes-rangel2020}.

In HH~900, the CO outflow morphology connects smoothly with the ionized outflow traced by H$\alpha$ and \brg\ that is seen outside the globule. 
Near-IR [Fe~{\sc ii}] emission traces the fast, collimated jet that bisects the wide-angle ionized outflow. 
However, this component is first detected $\gtrsim$1.5\arcsec\ from the edges of the globule. 
Non-instantaneous dissociation of the molecular outflow provides one explanation for this gap.  
In this region, all UV photons are consumed ionizing the skin of the outflow (\brg; H$\alpha$) and dissociating molecules just inside it (\mh; [C~{\sc i}]), exhausting the high energy photons before they can reach the core and ionize Fe (first ionization requires 7.6~eV). 
Essentially, slices along the outflow at different distances from the globule trace the development and progression of a photodissociation region (PDR) through the column of outflowing gas (see Figure~\ref{fig:cartoon}).

In a less punishing environment than Carina, HH~900 would look like a traditional molecular outflow with CO tracing more of its extent. 
The HH~900~YSO remains deeply embedded in the globule with the only detection to date in the millimeter continuum with ALMA. 
This suggests that the source evolutionary stage is in the Class~0/I regime, consistent with HH~900 being in the most active outflow phase (we discuss the mass-loss rates in Section~\ref{ss:h2_mdot}).

One of the outstanding challenges for understanding the local and large-scale impact of protostellar jets and outflows is obtaining a full mass census.  Typically, only one component is visible: 
the underlying (atomic) jet is unseen in molecular outflows while optical and near-IR images reveal collimated atomic jets that have largely escaped their natal clouds and may no longer be surrounded by a molecular outflow. 
External illumination reveals all of these components in the HH~900 jet+outflow. 
The system is younger than other well-studied examples of externally irradiated jets \citep[e.g., in Orion,][]{bal01,bal06,kirwan2023}, capturing a consequential but not well understood outflow phase.  

External irradiation in the H~{\sc ii} region alters the observable diagnostics but provides an extraordinary opportunity because it also illuminates the entire body of the jet+outflow. 
Mass-loss rates can therefore be estimated using the well-understood physics of photoionized gas instead of non-linear and time-dependent shock models. 
Following \citet{bal06}, we can use the well-studied environment to determine the dissociation time, and therefore the mass-loss rate, of the dissociating molecular outflow. 

\subsection{Mass-loss rate}\label{ss:h2_mdot}

We make a simple estimate of the photodissociation rate of molecular hydrogen entrained in the outflow. The free space photodissociation rate is $D_0 = 5.18\times10^{-11}\chi$\,s$^{-1}$ \citep{2014ApJ...790...10S, 2020CmPhy...3...32B} where $\chi$ is the Draine UV radiation field strength \citep{1978ApJS...36..595D}. Ignoring shielding, the photodissociation rate of \mh\ at the surface of a slab irradiated from one side (the outflow in our case) is 
\begin{equation}
   R_{\mathrm{phot}} = \frac{1}{2} \chi 5.18 \times 10^{-11}~\mathrm{s}^{-1}.
\end{equation}
From \citet{smith2006_energy}, the FUV luminosity of Tr16 is  
log(L$_{\mathrm{FUV}}$) $= 6.79$~L$_{\odot}$. 
We convert this to the local flux assuming the median distance to the OB stars in Tr16 from \citet{alexander2016}.  
This gives an incident Draine UV radiation field of $\chi=10^3$, corresponding to a dissociation timescale of $\sim$1.2 years. 
For an outflow velocity of 30\,km\,s$^{-1}$, this gives a distance of $\sim$8\,au for the extent of the \mh\ outflow. 
This is $\sim$3 orders of magnitude smaller than the observed \mh\ extent of HH~900 ($\sim$5000\,au, see Section~\ref{ss:morphology}).  

This simple estimate ignores the effect of shielding. 
However, \mh\ can be shielded by dust \citep[which may also be present in the outflow][]{smith2010,reiter2019_tadpole,reiter2020_tadpole} or it can self-shield, provided a sufficiently large \mh\ column \citep{2014ApJ...790...10S}. 
Including self-shielding, the dissociation rate becomes 
\begin{equation}
   R_{\mathrm{phot}} = \frac{1}{2} \chi 5.18 \times 10^{-11} f_{\mathrm{shield}}(N_{\mh})\exp\left(-\tau_g\right)\mathrm{s}^{-1}.
\end{equation}
where $\tau_g = \sigma_g N_{\mh}$ is the dust column optical depth and $f_{\mathrm{shield}}(N_{\mh})$ is the H$_2$ self-shielding parameter; both depend on the H$_2$ column $N_{\mh}$. We follow \cite{2020CmPhy...3...32B} and adopt $\sigma_g = 1.9\times10^{-21}$cm$^2$. For $f_{\mathrm{shield}}(N_{\mh})$ we use the \cite{1996ApJ...468..269D} fit
%
\begin{multline}
    f_{\mathrm{shield}}(N_{\mh}) = \frac{0.965}{\left(1+x/b_5\right)^2} + \frac{0.035}{\left(1+x\right)^{1/2}} \\
    \times \exp\left[-8.5\times10^{-4}\left(1+x\right)^{1/2}\right]    
\end{multline}
%
where $x = N_{\mh}/5\times10^{14}$cm$^{-2}$ and $b_5=2/10^5$cm\,s$^{-1}$. 
This gives an \mh\ photodissociation timescale that depends only on the \mh\ column for a given $\chi$. 
We solve this expression for the \mh\ column that enables the H$_2$ outflow to extend $\sim 5 \times 10^3$\,au with a flow velocity of $\sim30$\,km\,s$^{-1}$. 
We calculate the photodissociation timescale $t_{\mathrm{phot}} = 1/R_{\mathrm{phot}}$ and multiply by a characteristic speed (30~\kms) to obtain the maximum observable \mh\ extent.

\begin{figure}
    \centering
    \includegraphics[width=\columnwidth]{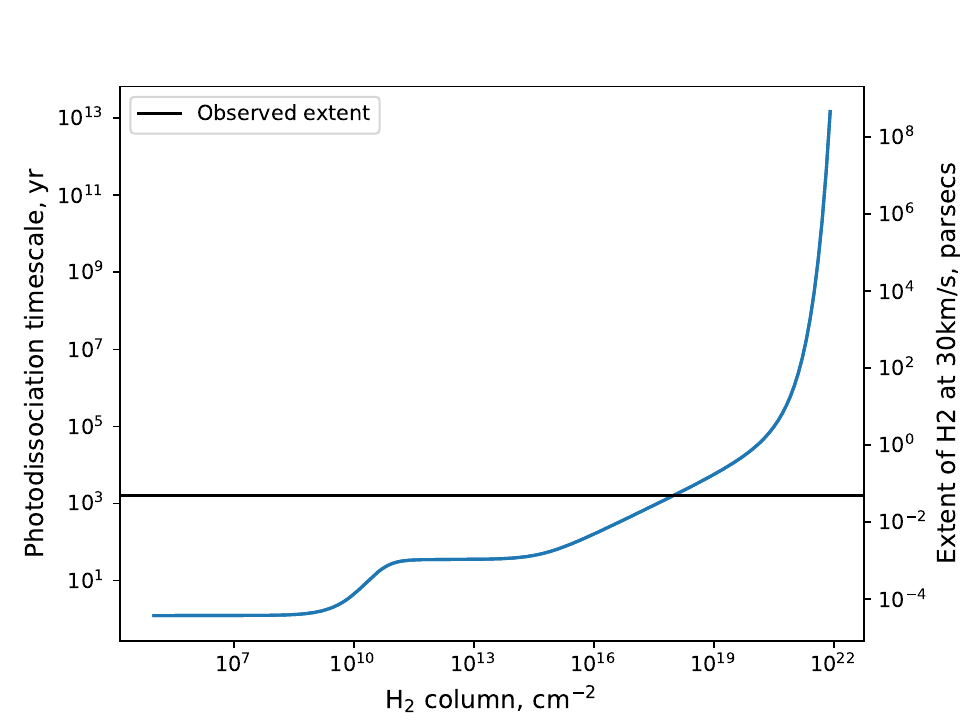}
    \caption{The photodissociation timescale as a function of \mh\ column (left axis). This is also presented in terms of the extent of molecular H$_2$ expected for a column travelling at 30\,km\,s$^{-1}$ (right axis). This is calculated by multiplying the photodissociation timescale $t_{\mathrm{phot}} = 1/R_{\mathrm{phot}}$ by the typical flow speed. The horizontal black line is the observed extent of \mh\ in the outflow.}
    \label{fig:H2extent}
\end{figure}

Results from this simple analysis are shown in Figure~\ref{fig:H2extent}. 
We estimate that an \mh\ column of $\sim 1 \times 10^{18}$~cm$^{-2}$ is required to reproduce the observed extent of the outflow. 
Assuming the outflow is a cylinder, we convert this to a volume density by dividing by the observed width of the outflow. 
For an \mh\ column density of $\sim 10^{18}$~cm$^{-2}$ and a jet width of $2r_{\mathrm{out}}\sim$1.5\arcsec\ (3450~au), the \mh\ volume density is $n_{\mh}\sim 20$~cm$^{-3}$. 
We compute the mass-loss rate of the \mh\ outflow assuming a constant density and radius in a cylindrical jet 
\begin{equation}
\dot{M}_{\mh}= n_{\mh} m_{\mh} \pi r_{\mathrm{out}}^2 v_{\mathrm{out}}     
\end{equation}
where $m_{\mh}$ is the mass of molecular hydrogen and $v_{\mathrm{out}} = 30$~\kms\ is the characteristic velocity of the outflow. 
From this, we estimate  
$\dot{M}_{\mh} \approx 1.6 \times 10^{-9}$~M$_{\odot}$~yr$^{-1}$.

This simple estimate yields a mass-loss rate two orders of magnitude smaller than estimated for the ionized component, $\dot{M}_{\mathrm{H}\alpha} \approx 5.7 \times 10^{-7}$~M$_{\odot}$~yr$^{-1}$ \citep{smith2010}. 
The mass-loss rate of the molecular outflow inside the globule is an order of magnitude higher than the ionized component with an average 
$\dot{M}_{\mathrm{CO}} \approx 3.5 \times 10^{-6}$~M$_{\odot}$~yr$^{-1}$ \citep{reiter2020_tadpole}. 
The highest mass-loss rate is measure in the low-ionization jet core with 
$\dot{M}_{\mathrm{[Fe II]}} \approx 1.7 \times 10^{-5}$~M$_{\odot}$~yr$^{-1}$ \citep{reiter2016}.

The large discrepancy between the mass-loss rates in the molecular outflow measured inside and outside the globule is not entirely surprising. 
We ignore any dust shielding in the outflow 
despite indirect evidence for dust in the outflow. 
In addition, we argue that \mh\ emission traces partially dissociated molecular gas. 
The column of \textit{neutral} material tracing the brief phase between dissociation and ionization should increase once the outflow is in the H~{\sc ii} region. 
Near the edge of the globule where the outflow first emerges into the H~{\sc ii} region, we expect that the sum of the molecular (\mh), neutral (C~{\sc i}?), and ionized (H$\alpha$ or \brg) components to be comparable to the CO mass-loss rate measured inside the globule. 
This implies that there is a large column of neutral material in the outflow. 
Future observations of neutral gas tracers like C~{\sc i} with ALMA may reveal the mass and extent of this neutral component.

Finally, we note that the low volume density we estimate in \mh\ is consistent with photoexcited gas. 
\citet{black1987} predict that, in steady state, the boundary layer should have $n(\mh) << n(\mathrm{H})$. 
We derive an $n(\mh)$ that is 13$\times$ smaller than the $n_e$ reported by \citet{smith2010}, consistent with this prediction.

\begin{figure}
    \centering
    \includegraphics[width=\columnwidth]{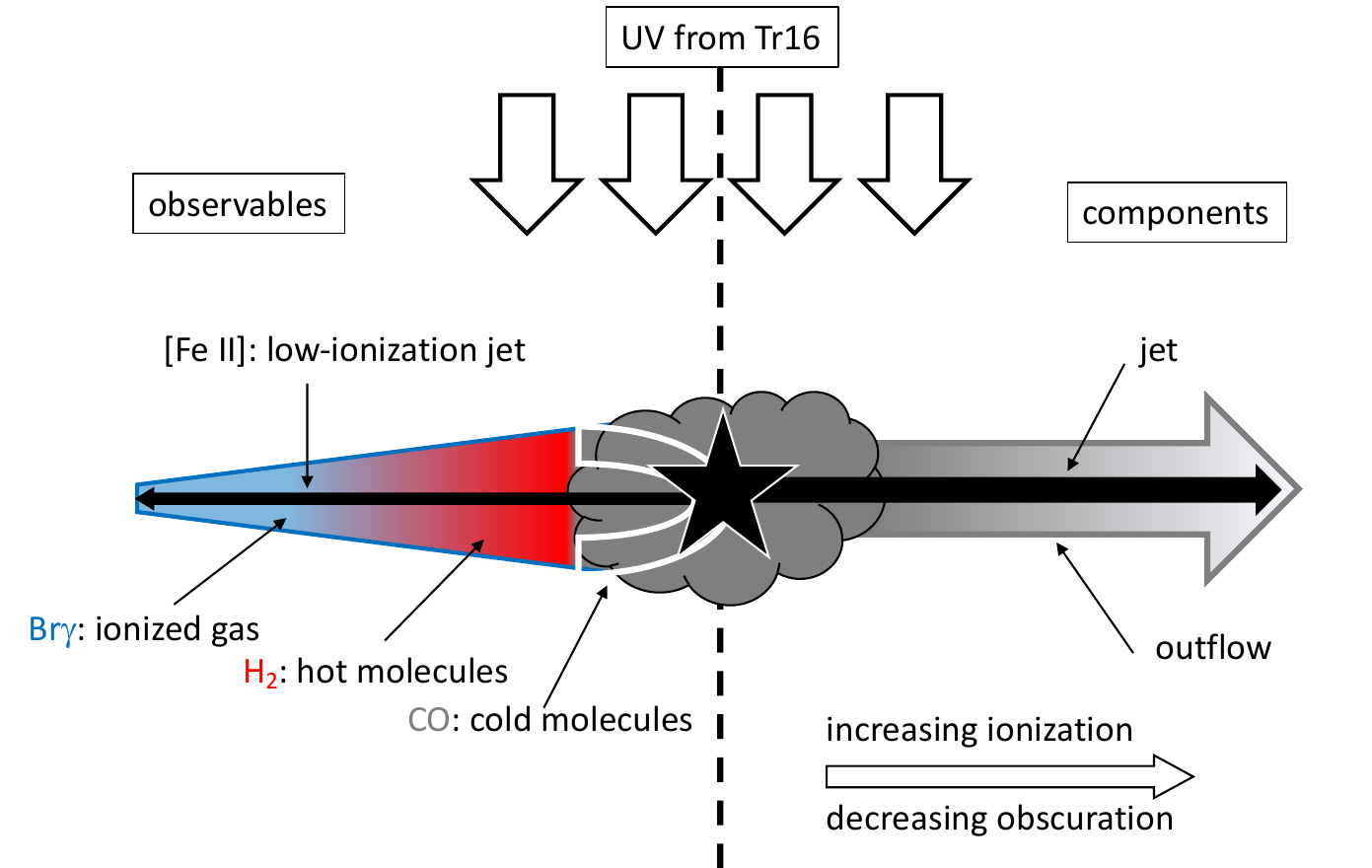}
    \caption{Cartoon depiction of the jet observables (left) and the structure of the jet+outflow (right). Cold molecules only survive in the protection of the globule (CO; gray). In the H~{\sc ii} region, molecules are hot (\mh; red) before they are completely dissociated leaving the ionized outflow (\brg; blue) surrounding the underlying jet ([Fe~{\sc ii}]; black). }
    \label{fig:cartoon}
\end{figure}

\section{Conclusions}\label{s:conclusions}

In H~{\sc ii} regions, jets and outflows that emerge from their natal clouds will be externally illuminated, dissociated, and ionized by UV photons from nearby high-mass stars. 
In this paper, we present new near-IR integral field unit spectroscopy from ERIS/SPIFFIER that provides the first clear evidence that  HH~900 is a dissociating and evaporating molecular outflow.   
Extended \mh\ and \brg\ emission trace the wide-angle outflow as it smoothly extends from the edge of the CO outflow. 
\mh\ emission extends $\sim$2.2\arcsec\ (0.02~pc) from the globule edge, tracing the molecule survival time. 
The ionized outflow, traced by \brg, reveals the full extent of the outflow as it reaches $\sim$5.5\arcsec\ into the H~{\sc ii} region.

These new spatially and spectrally resolved observations of HH~900 allow us to perform three tests to confirm that HH~900 is an evaporating molecular outflow. 
First, we show that \mh\ and \brg\ trace the same morphology as long as molecules survive. 
Second, we show that both lines trace outflow-like kinematics. 
Velocities of both lines are modest where the outflow emerges from the globule, then \mh\ increases to speeds approaching the high velocity of the underlying jet seen in [Fe~{\sc ii}]. 
Velocities between each component differ by $\sim$10~\kms\ with [Fe~{\sc ii}] tracing the fastest velocities, \mh\ intermediate, and \brg\ the slowest, consistent with the layered velocity structure seen in other outflows. 
Third, diagnostic line ratios indicate a significant contribution from photoexcitation to the excitation in the outflow. 
Together, these tests provide strong evidence that an evaporating molecular outflow connects the cold (CO) outflow seen inside the globule with the hot ionized outflow (H$\alpha$, \brg) seen in the H~{\sc ii} region.

To the best of our knowledge, this is the first direct evidence for an evaporating molecular outflow. 
As such, these observations provide an excellent test case for models of irradiated jets and outflows \citep[i.e.,][]{estrella-trujillo2021}.  
Finally, near-IR observations from the \emph{James Webb Space Telescope} (JWST) will soon be revealing jets and outflows in many high-mass star-forming regions \citep[e.g.,][]{reiter2022_jwst}.
These externally irradiated sources will likely have more in common with objects like HH~900 than the well-studied jets seen in more local, quiescent regions.

\section*{Acknowledgements}
We would like to thank the referee, Dr. William Henney, for a prompt and thoughtful report that improved the quality of the manuscript. 
We would like to thank Lowell Tacconi-Garman for support with the preparation, reduction, and analysis of the new ERIS/SPIFFIER data. 
We wish to thank Nathan Smith for productive discussion about the velocity structure of HH~900.
TJH is funded by a Royal Society Dorothy Hodgkin Fellowship and UKRI ERC guarantee funding (EP/Y024710/1). 
CFM is funded by the European Union (ERC, WANDA, 101039452). 
DI is funded by the European Research Council (ERC) via the ERC Synergy Grant ECOGAL (grant 855130).  
Views and opinions expressed are however those of the author(s) only and do not necessarily reflect those of the European Union or the European Research Council Executive Agency. Neither the European Union nor the granting authority can be held responsible for them.

Based on observations collected at the European Southern Observatory under ESO programmes 110.257T.001 and 0101.C-0391(A).
This paper makes use of the following ALMA data: ADS/JAO.ALMA \#2016.1.01537.S.
ALMA is a partnership of ESO (representing its member states), NSF (USA) and NINS (Japan), together with NRC (Canada) and NSC and ASIAA (Taiwan) and KASI (Republic of Korea), in cooperation with the Republic of Chile. The Joint ALMA Observatory is operated by ESO, AUI/NRAO and NAOJ.
This work uses observations made with the NASA/ESA Hubble Space Telescope, obtained from the Data Archive at the Space Telescope Science Institute, which is operated by the Association of Universities for Research in Astronomy, Inc., under NASA contract NAS 5-26555.
The HST observations are associated with GO~13390 and GO~13391. 
This research made use of Astropy,\footnote{\href{http://www.astropy.org}{http://www.astropy.org}} a community-developed core Python package for Astronomy \citep{astropy:2013, astropy:2018}.
This research made use of APLpy, an open-source plotting package for Python \citep{robitaille2012}. 

\section*{Data Availability}

 The ERIS data that are presented in this paper are publicly available from the ESO archive\footnote{\href{https://archive.eso.org/cms.html}{https://archive.eso.org/cms.html}}. 
Archival MUSE data are also available from the ESO archive. 
The ALMA data used in this study are publicly available from the ALMA archive\footnote{\href{https://almascience.nrao.edu/aq/?result_view=observations}{https://almascience.nrao.edu/aq/?result\_view=observations}}. 
Data from \emph{HST} are publicly available via the MAST archive\footnote{\href{https://mast.stsci.edu/portal/Mashup/Clients/Mast/Portal.html}{https://mast.stsci.edu/portal/Mashup/Clients/Mast/Portal.html}}.



\bibliographystyle{mnras}
\bibliography{bibliography} 

\begin{thebibliography}{}
\makeatletter
\relax
\def\mn@urlcharsother{\let\do\@makeother \do\$\do\&\do\#\do\^\do\_\do\%\do\~}
\def\mn@doi{\begingroup\mn@urlcharsother \@ifnextchar [ {\mn@doi@}
  {\mn@doi@[]}}
\def\mn@doi@[#1]#2{\def\@tempa{#1}\ifx\@tempa\@empty \href
  {http://dx.doi.org/#2} {doi:#2}\else \href {http://dx.doi.org/#2} {#1}\fi
  \endgroup}
\def\mn@eprint#1#2{\mn@eprint@#1:#2::\@nil}
\def\mn@eprint@arXiv#1{\href {http://arxiv.org/abs/#1} {{\tt arXiv:#1}}}
\def\mn@eprint@dblp#1{\href {http://dblp.uni-trier.de/rec/bibtex/#1.xml}
  {dblp:#1}}
\def\mn@eprint@#1:#2:#3:#4\@nil{\def\@tempa {#1}\def\@tempb {#2}\def\@tempc
  {#3}\ifx \@tempc \@empty \let \@tempc \@tempb \let \@tempb \@tempa \fi \ifx
  \@tempb \@empty \def\@tempb {arXiv}\fi \@ifundefined
  {mn@eprint@\@tempb}{\@tempb:\@tempc}{\expandafter \expandafter \csname
  mn@eprint@\@tempb\endcsname \expandafter{\@tempc}}}

\bibitem[\protect\citeauthoryear{{Alexander}, {Hanes}, {Povich}  \&
  {McSwain}}{{Alexander} et~al.}{2016}]{alexander2016}
{Alexander} M.~J.,  {Hanes} R.~J.,  {Povich} M.~S.,   {McSwain} M.~V.,  2016,
  \mn@doi [\aj] {10.3847/0004-6256/152/6/190}, \href
  {http://adsabs.harvard.edu/abs/2016AJ....152..190A} {152, 190}

\bibitem[\protect\citeauthoryear{{Arce}, {Shepherd}, {Gueth}, {Lee},
  {Bachiller}, {Rosen}  \& {Beuther}}{{Arce} et~al.}{2007}]{arc07}
{Arce} H.~G.,  {Shepherd} D.,  {Gueth} F.,  {Lee} C.-F.,  {Bachiller} R.,
  {Rosen} A.,   {Beuther} H.,  2007, Protostars and Planets V, \href
  {http://adsabs.harvard.edu/abs/2007prpl.conf..245A} {pp 245--260}

\bibitem[\protect\citeauthoryear{{Arce}, {Mardones}, {Corder}, {Garay},
  {Noriega-Crespo}  \& {Raga}}{{Arce} et~al.}{2013}]{arc13}
{Arce} H.~G.,  {Mardones} D.,  {Corder} S.~A.,  {Garay} G.,  {Noriega-Crespo}
  A.,   {Raga} A.~C.,  2013, \mn@doi [\apj] {10.1088/0004-637X/774/1/39}, \href
  {http://adsabs.harvard.edu/abs/2013ApJ...774...39A} {774, 39}

\bibitem[\protect\citeauthoryear{{Astropy Collaboration} et~al.,}{{Astropy
  Collaboration} et~al.}{2013}]{astropy:2013}
{Astropy Collaboration} et~al., 2013, \mn@doi [aap]
  {10.1051/0004-6361/201322068}, \href
  {http://adsabs.harvard.edu/abs/2013A%26A...558A..33A} {558, A33}

\bibitem[\protect\citeauthoryear{{Bacciotti}, {Mundt}, {Ray}, {Eisl{\"o}ffel},
  {Solf}  \& {Camezind}}{{Bacciotti} et~al.}{2000}]{bac00}
{Bacciotti} F.,  {Mundt} R.,  {Ray} T.~P.,  {Eisl{\"o}ffel} J.,  {Solf} J.,
  {Camezind} M.,  2000, \mn@doi [\apjl] {10.1086/312745}, \href
  {http://adsabs.harvard.edu/abs/2000ApJ...537L..49B} {537, L49}

\bibitem[\protect\citeauthoryear{{Bally}}{{Bally}}{2016}]{bally2016}
{Bally} J.,  2016, \mn@doi [\araa] {10.1146/annurev-astro-081915-023341}, \href
  {https://ui.adsabs.harvard.edu/abs/2016ARA&A..54..491B} {54, 491}

\bibitem[\protect\citeauthoryear{{Bally} \& {Reipurth}}{{Bally} \&
  {Reipurth}}{2001}]{bal01}
{Bally} J.,  {Reipurth} B.,  2001, \mn@doi [\apj] {10.1086/318258}, \href
  {http://adsabs.harvard.edu/abs/2001ApJ...546..299B} {546, 299}

\bibitem[\protect\citeauthoryear{{Bally}, {Licht}, {Smith}  \&
  {Walawender}}{{Bally} et~al.}{2006a}]{bal06}
{Bally} J.,  {Licht} D.,  {Smith} N.,   {Walawender} J.,  2006a, \mn@doi [\aj]
  {10.1086/498265}, \href {http://adsabs.harvard.edu/abs/2006AJ....131..473B}
  {131, 473}

\bibitem[\protect\citeauthoryear{{Bally}, {Licht}, {Smith}  \&
  {Walawender}}{{Bally} et~al.}{2006b}]{bally2006}
{Bally} J.,  {Licht} D.,  {Smith} N.,   {Walawender} J.,  2006b, \mn@doi [\aj]
  {10.1086/498265}, \href
  {https://ui.adsabs.harvard.edu/abs/2006AJ....131..473B} {131, 473}

\bibitem[\protect\citeauthoryear{{Bialy}}{{Bialy}}{2020}]{2020CmPhy...3...32B}
{Bialy} S.,  2020, \mn@doi [Communications Physics]
  {10.1038/s42005-020-0293-7}, \href
  {https://ui.adsabs.harvard.edu/abs/2020CmPhy...3...32B} {3, 32}

\bibitem[\protect\citeauthoryear{{Black} \& {van Dishoeck}}{{Black} \& {van
  Dishoeck}}{1987}]{black1987}
{Black} J.~H.,  {van Dishoeck} E.~F.,  1987, \mn@doi [\apj] {10.1086/165740},
  \href {https://ui.adsabs.harvard.edu/abs/1987ApJ...322..412B} {322, 412}

\bibitem[\protect\citeauthoryear{{Calvet}, {Patino}, {Magris}  \&
  {D'Alessio}}{{Calvet} et~al.}{1991}]{calvet1991}
{Calvet} N.,  {Patino} A.,  {Magris} G.~C.,   {D'Alessio} P.,  1991, \mn@doi
  [\apj] {10.1086/170618}, \href
  {http://adsabs.harvard.edu/abs/1991ApJ...380..617C} {380, 617}

\bibitem[\protect\citeauthoryear{{Carr}}{{Carr}}{1989}]{carr1989}
{Carr} J.~S.,  1989, \mn@doi [\apj] {10.1086/167927}, \href
  {http://adsabs.harvard.edu/abs/1989ApJ...345..522C} {345, 522}

\bibitem[\protect\citeauthoryear{Chang \& Deming}{Chang \&
  Deming}{1996}]{chang1996observation}
Chang E.~S.,  Deming D.,  1996, Solar Physics, 165, 257

\bibitem[\protect\citeauthoryear{{Coffey}, {Bacciotti}  \& {Podio}}{{Coffey}
  et~al.}{2008}]{coffey2008}
{Coffey} D.,  {Bacciotti} F.,   {Podio} L.,  2008, \mn@doi [\apj]
  {10.1086/592343}, \href
  {https://ui.adsabs.harvard.edu/abs/2008ApJ...689.1112C} {689, 1112}

\bibitem[\protect\citeauthoryear{{Cortes-Rangel}, {Zapata}, {Toal{\'a}}, {Ho},
  {Takahashi}, {Mesa-Delgado}  \& {Masqu{\'e}}}{{Cortes-Rangel}
  et~al.}{2020}]{cortes-rangel2020}
{Cortes-Rangel} G.,  {Zapata} L.~A.,  {Toal{\'a}} J.~A.,  {Ho} P. T.~P.,
  {Takahashi} S.,  {Mesa-Delgado} A.,   {Masqu{\'e}} J.~M.,  2020, \mn@doi
  [\aj] {10.3847/1538-3881/ab6295}, \href
  {https://ui.adsabs.harvard.edu/abs/2020AJ....159...62C} {159, 62}

\bibitem[\protect\citeauthoryear{{Davies} et~al.,}{{Davies}
  et~al.}{2018}]{davies2018}
{Davies} R.,  et~al., 2018, in {Evans} C.~J.,  {Simard} L.,   {Takami} H.,
  eds,  Society of Photo-Optical Instrumentation Engineers (SPIE) Conference
  Series Vol. 10702, Ground-based and Airborne Instrumentation for Astronomy
  VII. p. 1070209 (\mn@eprint {arXiv} {1807.05089}),
  \mn@doi{10.1117/12.2311480}

\bibitem[\protect\citeauthoryear{{Davies} et~al.,}{{Davies}
  et~al.}{2023}]{davies2023}
{Davies} R.,  et~al., 2023, \mn@doi [\aap] {10.1051/0004-6361/202346559}, \href
  {https://ui.adsabs.harvard.edu/abs/2023A&A...674A.207D} {674, A207}

\bibitem[\protect\citeauthoryear{{Draine}}{{Draine}}{1978}]{1978ApJS...36..595D}
{Draine} B.~T.,  1978, \mn@doi [\apjs] {10.1086/190513}, \href
  {https://ui.adsabs.harvard.edu/abs/1978ApJS...36..595D} {36, 595}

\bibitem[\protect\citeauthoryear{{Draine} \& {Bertoldi}}{{Draine} \&
  {Bertoldi}}{1996}]{1996ApJ...468..269D}
{Draine} B.~T.,  {Bertoldi} F.,  1996, \mn@doi [\apj] {10.1086/177689}, \href
  {https://ui.adsabs.harvard.edu/abs/1996ApJ...468..269D} {468, 269}

\bibitem[\protect\citeauthoryear{{Erkal} et~al.,}{{Erkal}
  et~al.}{2021}]{erkal2021}
{Erkal} J.,  et~al., 2021, \mn@doi [\apj] {10.3847/1538-4357/ac06c5}, \href
  {https://ui.adsabs.harvard.edu/abs/2021ApJ...919...23E} {919, 23}

\bibitem[\protect\citeauthoryear{{Escalante}, {Sternberg}  \&
  {Dalgarno}}{{Escalante} et~al.}{1991}]{escalante1991}
{Escalante} V.,  {Sternberg} A.,   {Dalgarno} A.,  1991, \mn@doi [\apj]
  {10.1086/170225}, \href
  {https://ui.adsabs.harvard.edu/abs/1991ApJ...375..630E} {375, 630}

\bibitem[\protect\citeauthoryear{{Estrella-Trujillo}, {Vel{\'a}zquez}, {Raga}
  \& {Esquivel}}{{Estrella-Trujillo} et~al.}{2021}]{estrella-trujillo2021}
{Estrella-Trujillo} D.,  {Vel{\'a}zquez} P.~F.,  {Raga} A.~C.,   {Esquivel} A.,
   2021, \mn@doi [\apj] {10.3847/1538-4357/ac1122}, \href
  {https://ui.adsabs.harvard.edu/abs/2021ApJ...918...75E} {918, 75}

\bibitem[\protect\citeauthoryear{{Fairlamb}, {Oudmaijer}, {Mendigutia}, {Ilee}
  \& {van den Ancker}}{{Fairlamb} et~al.}{2017}]{fairlamb2017}
{Fairlamb} J.~R.,  {Oudmaijer} R.~D.,  {Mendigutia} I.,  {Ilee} J.~D.,   {van
  den Ancker} M.~E.,  2017, \mn@doi [\mnras] {10.1093/mnras/stw2643}, \href
  {https://ui.adsabs.harvard.edu/abs/2017MNRAS.464.4721F} {464, 4721}

\bibitem[\protect\citeauthoryear{{Freudling}, {Romaniello}, {Bramich},
  {Ballester}, {Forchi}, {Garc{\'{\i}}a-Dabl{\'o}}, {Moehler}  \&
  {Neeser}}{{Freudling} et~al.}{2013}]{reflex}
{Freudling} W.,  {Romaniello} M.,  {Bramich} D.~M.,  {Ballester} P.,  {Forchi}
  V.,  {Garc{\'{\i}}a-Dabl{\'o}} C.~E.,  {Moehler} S.,   {Neeser} M.~J.,  2013,
  \mn@doi [\aap] {10.1051/0004-6361/201322494}, \href
  {http://adsabs.harvard.edu/abs/2013A%26A...559A..96F} {559, A96}

\bibitem[\protect\citeauthoryear{{G{\"o}ppl} \& {Preibisch}}{{G{\"o}ppl} \&
  {Preibisch}}{2022}]{goeppl2022}
{G{\"o}ppl} C.,  {Preibisch} T.,  2022, \mn@doi [\aap]
  {10.1051/0004-6361/202142576}, \href
  {https://ui.adsabs.harvard.edu/abs/2022A&A...660A..11G} {660, A11}

\bibitem[\protect\citeauthoryear{{Hartigan}, {Heathcote}, {Morse}, {Reipurth}
  \& {Bally}}{{Hartigan} et~al.}{2005}]{hartigan2005}
{Hartigan} P.,  {Heathcote} S.,  {Morse} J.~A.,  {Reipurth} B.,   {Bally} J.,
  2005, \mn@doi [\aj] {10.1086/491673}, \href
  {https://ui.adsabs.harvard.edu/abs/2005AJ....130.2197H} {130, 2197}

\bibitem[\protect\citeauthoryear{{Hartigan} et~al.,}{{Hartigan}
  et~al.}{2011}]{hartigan2011}
{Hartigan} P.,  et~al., 2011, \mn@doi [\apj] {10.1088/0004-637X/736/1/29},
  \href {https://ui.adsabs.harvard.edu/abs/2011ApJ...736...29H} {736, 29}

\bibitem[\protect\citeauthoryear{{Hartigan}, {Reiter}, {Smith}  \&
  {Bally}}{{Hartigan} et~al.}{2015}]{hartigan2015}
{Hartigan} P.,  {Reiter} M.,  {Smith} N.,   {Bally} J.,  2015, \mn@doi [\aj]
  {10.1088/0004-6256/149/3/101}, \href
  {https://ui.adsabs.harvard.edu/abs/2015AJ....149..101H} {149, 101}

\bibitem[\protect\citeauthoryear{{Heathcote}, {Morse}, {Hartigan}, {Reipurth},
  {Schwartz}, {Bally}  \& {Stone}}{{Heathcote} et~al.}{1996}]{hea96}
{Heathcote} S.,  {Morse} J.~A.,  {Hartigan} P.,  {Reipurth} B.,  {Schwartz}
  R.~D.,  {Bally} J.,   {Stone} J.~M.,  1996, \mn@doi [\aj] {10.1086/118085},
  \href {http://adsabs.harvard.edu/abs/1996AJ....112.1141H} {112, 1141}

\bibitem[\protect\citeauthoryear{{Kiminki} \& {Smith}}{{Kiminki} \&
  {Smith}}{2018}]{kiminki2018}
{Kiminki} M.~M.,  {Smith} N.,  2018, \mn@doi [\mnras] {10.1093/mnras/sty748},
  \href {http://adsabs.harvard.edu/abs/2018MNRAS.477.2068K} {477, 2068}

\bibitem[\protect\citeauthoryear{{Kirwan} et~al.,}{{Kirwan}
  et~al.}{2023}]{kirwan2023}
{Kirwan} A.,  et~al., 2023, \mn@doi [\aap] {10.1051/0004-6361/202245428}, \href
  {https://ui.adsabs.harvard.edu/abs/2023A&A...673A.166K} {673, A166}

\bibitem[\protect\citeauthoryear{{Lavalley-Fouquet}, {Cabrit}  \&
  {Dougados}}{{Lavalley-Fouquet} et~al.}{2000}]{lavalley-fouquet2000}
{Lavalley-Fouquet} C.,  {Cabrit} S.,   {Dougados} C.,  2000, \aap, \href
  {https://ui.adsabs.harvard.edu/abs/2000A&A...356L..41L} {356, L41}

\bibitem[\protect\citeauthoryear{{Levenson} et~al.,}{{Levenson}
  et~al.}{2000}]{levenson2000}
{Levenson} N.~A.,  et~al., 2000, \mn@doi [\apjl] {10.1086/312601}, \href
  {https://ui.adsabs.harvard.edu/abs/2000ApJ...533L..53L} {533, L53}

\bibitem[\protect\citeauthoryear{{Noriega-Crespo} et~al.,}{{Noriega-Crespo}
  et~al.}{2004}]{nor04}
{Noriega-Crespo} A.,  et~al., 2004, \mn@doi [\apjs] {10.1086/422819}, \href
  {http://adsabs.harvard.edu/abs/2004ApJS..154..352N} {154, 352}

\bibitem[\protect\citeauthoryear{{Osterbrock} \& {Ferland}}{{Osterbrock} \&
  {Ferland}}{2006}]{osterbrock2006}
{Osterbrock} D.~E.,  {Ferland} G.~J.,  2006, {Astrophysics of gaseous nebulae
  and active galactic nuclei}

\bibitem[\protect\citeauthoryear{{Povich} et~al.,}{{Povich}
  et~al.}{2011}]{povich2011}
{Povich} M.~S.,  et~al., 2011, \mn@doi [\apjs] {10.1088/0067-0049/194/1/14},
  \href {http://adsabs.harvard.edu/abs/2011ApJS..194...14P} {194, 14}

\bibitem[\protect\citeauthoryear{{Price-Whelan} et~al.,}{{Price-Whelan}
  et~al.}{2018}]{astropy:2018}
{Price-Whelan} A.~M.,  et~al., 2018, \mn@doi [aj] {10.3847/1538-3881/aabc4f},
  \href {https://ui.adsabs.harvard.edu/#abs/2018AJ....156..123T} {156, 123}

\bibitem[\protect\citeauthoryear{{Pyo} et~al.,}{{Pyo} et~al.}{2003}]{pyo03}
{Pyo} T.-S.,  et~al., 2003, \mn@doi [\apj] {10.1086/374966}, \href
  {http://adsabs.harvard.edu/abs/2003ApJ...590..340P} {590, 340}

\bibitem[\protect\citeauthoryear{{Rebolledo} et~al.,}{{Rebolledo}
  et~al.}{2016}]{rebolledo2016}
{Rebolledo} D.,  et~al., 2016, \mn@doi [\mnras] {10.1093/mnras/stv2776}, \href
  {http://adsabs.harvard.edu/abs/2016MNRAS.456.2406R} {456, 2406}

\bibitem[\protect\citeauthoryear{{Reipurth}, {Bally}, {Fesen}  \&
  {Devine}}{{Reipurth} et~al.}{1998}]{rei98}
{Reipurth} B.,  {Bally} J.,  {Fesen} R.~A.,   {Devine} D.,  1998, \mn@doi
  [\nat] {10.1038/24562}, \href
  {http://adsabs.harvard.edu/abs/1998Natur.396..343R} {396, 343}

\bibitem[\protect\citeauthoryear{{Reiter}, {Smith}, {Kiminki}, {Bally}  \&
  {Anderson}}{{Reiter} et~al.}{2015a}]{reiter2015_hh900}
{Reiter} M.,  {Smith} N.,  {Kiminki} M.~M.,  {Bally} J.,   {Anderson} J.,
  2015a, \mn@doi [\mnras] {10.1093/mnras/stu177}, \href
  {http://adsabs.harvard.edu/abs/2015arXiv150106564R} {448, 3429}

\bibitem[\protect\citeauthoryear{{Reiter}, {Smith}, {Kiminki}  \&
  {Bally}}{{Reiter} et~al.}{2015b}]{reiter2015_hh666}
{Reiter} M.,  {Smith} N.,  {Kiminki} M.~M.,   {Bally} J.,  2015b, \mn@doi
  [\mnras] {10.1093/mnras/stv634}, \href
  {http://adsabs.harvard.edu/abs/2015MNRAS.450..564R} {450, 564}

\bibitem[\protect\citeauthoryear{{Reiter}, {Smith}  \& {Bally}}{{Reiter}
  et~al.}{2016}]{reiter2016}
{Reiter} M.,  {Smith} N.,   {Bally} J.,  2016, \mn@doi [\mnras]
  {10.1093/mnras/stw2296}, \href
  {http://adsabs.harvard.edu/abs/2016MNRAS.463.4344R} {463, 4344}

\bibitem[\protect\citeauthoryear{{Reiter}, {McLeod}, {Klaassen}, {Guzm{\'a}n},
  {Dale}, {Mottram}  \& {Garay}}{{Reiter} et~al.}{2019}]{reiter2019_tadpole}
{Reiter} M.,  {McLeod} A.~F.,  {Klaassen} P.~D.,  {Guzm{\'a}n} A.~E.,  {Dale}
  J.~E.,  {Mottram} J.~C.,   {Garay} G.,  2019, \mn@doi [\mnras]
  {10.1093/mnras/stz2752}, \href
  {https://ui.adsabs.harvard.edu/abs/2019MNRAS.490.2056R} {490, 2056}

\bibitem[\protect\citeauthoryear{{Reiter}, {Guzm{\'a}n}, {Haworth}, {Klaassen},
  {McLeod}, {Garay}  \& {Mottram}}{{Reiter} et~al.}{2020a}]{reiter2020_tadpole}
{Reiter} M.,  {Guzm{\'a}n} A.~E.,  {Haworth} T.~J.,  {Klaassen} P.~D.,
  {McLeod} A.~F.,  {Garay} G.,   {Mottram} J.~C.,  2020a, \mn@doi [\mnras]
  {10.1093/mnras/staa1504}, \href
  {https://ui.adsabs.harvard.edu/abs/2020MNRAS.496..394R} {496, 394}

\bibitem[\protect\citeauthoryear{{Reiter}, {Haworth}, {Guzm{\'a}n}, {Klaassen},
  {McLeod}  \& {Garay}}{{Reiter} et~al.}{2020b}]{reiter2020_tadpole_comp}
{Reiter} M.,  {Haworth} T.~J.,  {Guzm{\'a}n} A.~E.,  {Klaassen} P.~D.,
  {McLeod} A.~F.,   {Garay} G.,  2020b, \mn@doi [\mnras]
  {10.1093/mnras/staa2156}, \href
  {https://ui.adsabs.harvard.edu/abs/2020MNRAS.497.3351R} {497, 3351}

\bibitem[\protect\citeauthoryear{{Reiter}, {Morse}, {Smith}, {Haworth}, {Kuhn}
  \& {Klaassen}}{{Reiter} et~al.}{2022}]{reiter2022_jwst}
{Reiter} M.,  {Morse} J.~A.,  {Smith} N.,  {Haworth} T.~J.,  {Kuhn} M.~A.,
  {Klaassen} P.~D.,  2022, \mn@doi [\mnras] {10.1093/mnras/stac2820}, \href
  {https://ui.adsabs.harvard.edu/abs/2022MNRAS.517.5382R} {517, 5382}

\bibitem[\protect\citeauthoryear{{Robitaille} \& {Bressert}}{{Robitaille} \&
  {Bressert}}{2012}]{robitaille2012}
{Robitaille} T.,  {Bressert} E.,  2012, {APLpy: Astronomical Plotting Library
  in Python}, Astrophysics Source Code Library (\mn@eprint {ascl} {1208.017})

\bibitem[\protect\citeauthoryear{{Shull} \& {Hollenbach}}{{Shull} \&
  {Hollenbach}}{1978}]{shull1978}
{Shull} J.~M.,  {Hollenbach} D.~J.,  1978, \mn@doi [\apj] {10.1086/155934},
  \href {https://ui.adsabs.harvard.edu/abs/1978ApJ...220..525S} {220, 525}

\bibitem[\protect\citeauthoryear{{Smith}}{{Smith}}{2006}]{smith2006_energy}
{Smith} N.,  2006, \mn@doi [\mnras] {10.1111/j.1365-2966.2006.10007.x}, \href
  {http://adsabs.harvard.edu/abs/2006MNRAS.367..763S} {367, 763}

\bibitem[\protect\citeauthoryear{{Smith}, {Bally}  \& {Brooks}}{{Smith}
  et~al.}{2004}]{smith2004_hh666}
{Smith} N.,  {Bally} J.,   {Brooks} K.~J.,  2004, \mn@doi [\aj]
  {10.1086/383291}, \href {http://adsabs.harvard.edu/abs/2004AJ....127.2793S}
  {127, 2793}

\bibitem[\protect\citeauthoryear{{Smith}, {Bally}  \& {Walborn}}{{Smith}
  et~al.}{2010}]{smith2010}
{Smith} N.,  {Bally} J.,   {Walborn} N.~R.,  2010, \mn@doi [\mnras]
  {10.1111/j.1365-2966.2010.16520.x}, \href
  {http://adsabs.harvard.edu/abs/2010MNRAS.405.1153S} {405, 1153}

\bibitem[\protect\citeauthoryear{{Stanke}, {McCaughrean}  \&
  {Zinnecker}}{{Stanke} et~al.}{1999}]{stanke1999}
{Stanke} T.,  {McCaughrean} M.~J.,   {Zinnecker} H.,  1999, \aap, \href
  {https://ui.adsabs.harvard.edu/abs/1999A&A...350L..43S} {350, L43}

\bibitem[\protect\citeauthoryear{{Sternberg}, {Le Petit}, {Roueff}  \& {Le
  Bourlot}}{{Sternberg} et~al.}{2014}]{2014ApJ...790...10S}
{Sternberg} A.,  {Le Petit} F.,  {Roueff} E.,   {Le Bourlot} J.,  2014, \mn@doi
  [\apj] {10.1088/0004-637X/790/1/10}, \href
  {https://ui.adsabs.harvard.edu/abs/2014ApJ...790...10S} {790, 10}

\bibitem[\protect\citeauthoryear{{Yeh}, {Seaquist}, {Matzner}  \&
  {Pellegrini}}{{Yeh} et~al.}{2015}]{yeh2015}
{Yeh} S.~C.~C.,  {Seaquist} E.~R.,  {Matzner} C.~D.,   {Pellegrini} E.~W.,
  2015, \mn@doi [\apj] {10.1088/0004-637X/807/2/117}, \href
  {http://adsabs.harvard.edu/abs/2015ApJ...807..117Y} {807, 117}

\bibitem[\protect\citeauthoryear{{Zhang} et~al.,}{{Zhang}
  et~al.}{2016}]{zhang2016}
{Zhang} Y.,  et~al., 2016, \mn@doi [\apj] {10.3847/0004-637X/832/2/158}, \href
  {https://ui.adsabs.harvard.edu/abs/2016ApJ...832..158Z} {832, 158}

\bibitem[\protect\citeauthoryear{{Zinnecker}, {McCaughrean}  \&
  {Rayner}}{{Zinnecker} et~al.}{1998}]{zinnecker1998}
{Zinnecker} H.,  {McCaughrean} M.~J.,   {Rayner} J.~T.,  1998, \mn@doi [\nat]
  {10.1038/29716}, \href
  {https://ui.adsabs.harvard.edu/abs/1998Natur.394..862Z} {394, 862}

\makeatother
\end{thebibliography}


\appendix

\section{Reconciling HH~900 velocities}\label{s:vel_WTF}

The systematic velocity of HH~900 was not known when  \citet{reiter2015_hh900} analyzed the [Fe~{\sc ii}] emission from the jet. 
The derived velocity of $\approx 15$~\kms\ is much redder than Carina itself which has a heliocentric velocity of $\approx -8.4$~\kms\  \citep{rebolledo2016}. 
Re-examining the spectra in \citet{reiter2015_hh900}, we find that the estimated systematic velocity is almost exactly equal to the heliocentric velocity correction for the date of observation (15.75~\kms).  

We have remeasured the HH~900 jet velocities from the FIRE spectra presented in \citet{reiter2015_hh900} including the heliocentric velocity correction and the known velocity of the system \citep[see Section~\ref{s:data} and][]{reiter2020_tadpole}. 
This centers the [Fe~{\sc ii}] emission symmetrically around the source velocity, extending to $\sim \pm 30$~\kms. 

We note that the jet velocities measured from [Fe~{\sc ii}] emission in \citet{reiter2015_hh900} and \citet{reiter2019_tadpole} were both asymmetric with the blueshifted jet extending to $\sim -20$~\kms\ while the redshifted jet extended to $\sim 40$~\kms. 
This offset corresponds to the blueshift of $\sim 10$~\kms\ of HH~900 relative to Carina. 

\section{P-V slices}\label{s:slices}

\begin{figure}
    \centering
    \includegraphics[trim=-4mm 25mm 0mm 0mm,scale=0.595]{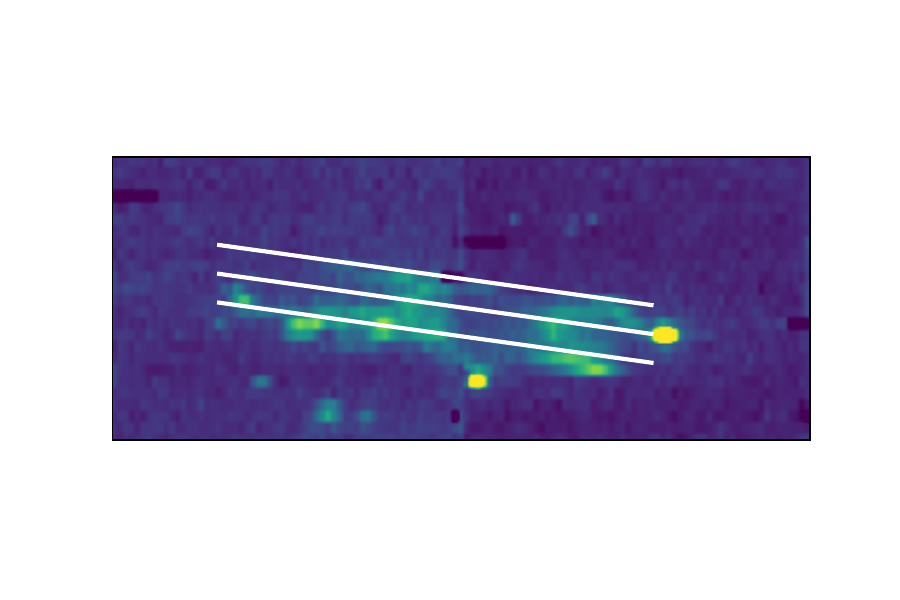}
    \includegraphics[width=\columnwidth,trim=0mm 5mm 0mm 0mm]{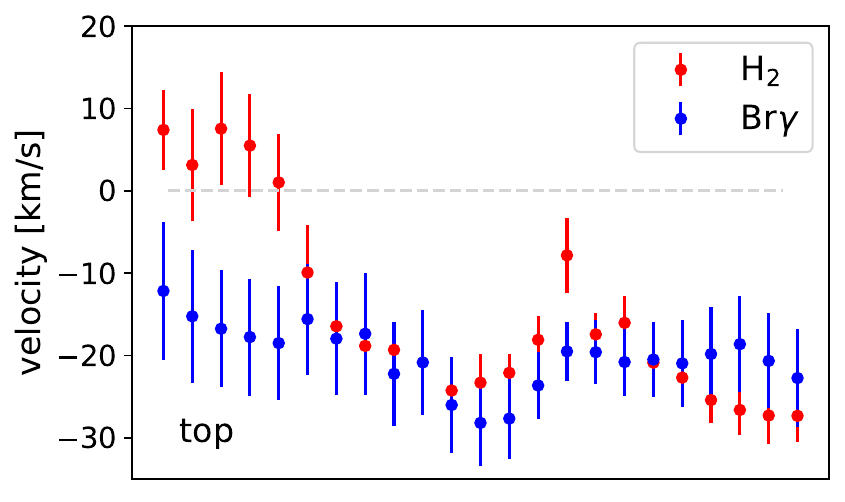}
    \includegraphics[width=\columnwidth,trim=0mm 5mm 0mm 0mm]{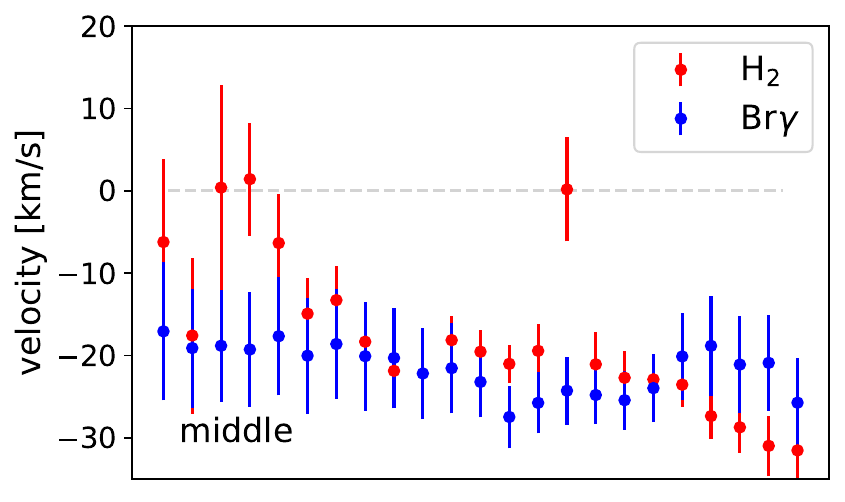}
    \includegraphics[width=\columnwidth,trim=0mm 5mm 0mm 0mm]{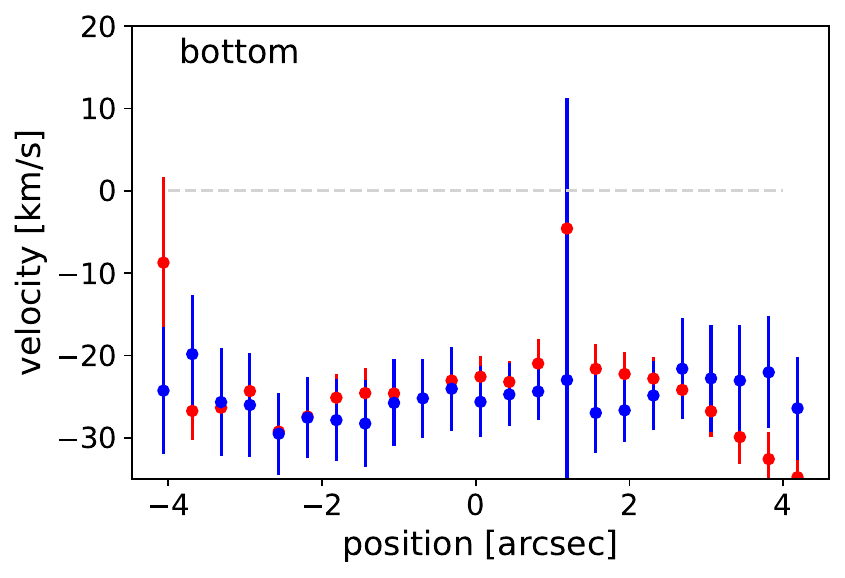}
    \caption{P-V slices extracted from the three positions in the tadpole shown in the top figure. The red-blue outflow velocities are most prominent in the top slice which is mininally affected by the \textit{blueshifted} tadpole tail. 
    Middle panel is the same as Figure~\ref{fig:pv}, reproduced here for convenience. 
    Bottom panel is the lowest slice with blueshifted emission from the tadpole tail clearly affecting the \mh\ velocities. }
    \label{fig:3_slices}
\end{figure}

We show three slices through the HH~900 jet+outflow in Figure~\ref{fig:3_slices}. 
As in Section~\ref{ss:velocities}, each slice is 5 spatial pixels wide. 
The slice through the top edge of HH~900 and the globule offer the most pristine measure of the outflow kinematics as this is unaffected by the tadpole tail. 
The middle P-V slice is the same as discussed in Section~\ref{ss:velocities}. 
The P-V slice along the bottom of the outflow and globule are clearly affected by the blueshifted tadpole tail which is brighter than the outflow in \mh. 
The tadpole globule is not resolved in \brg. 
Based on diffraction-limited H$\alpha$ images (see Figure~\ref{fig:context}), we expect the tail, like the globule, would be seen in silhouette, so does not affect the \brg\ velocities.

\begin{figure}
    \centering
    \includegraphics[trim=-4mm 25mm 0mm 0mm,scale=0.595]{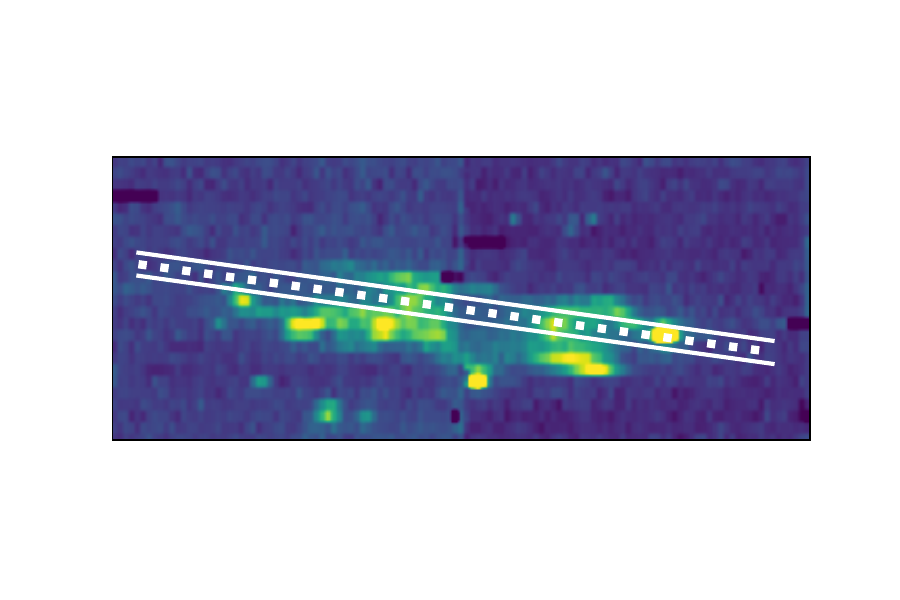}
    \includegraphics[width=\columnwidth,trim=0mm 5mm 0mm 0mm]{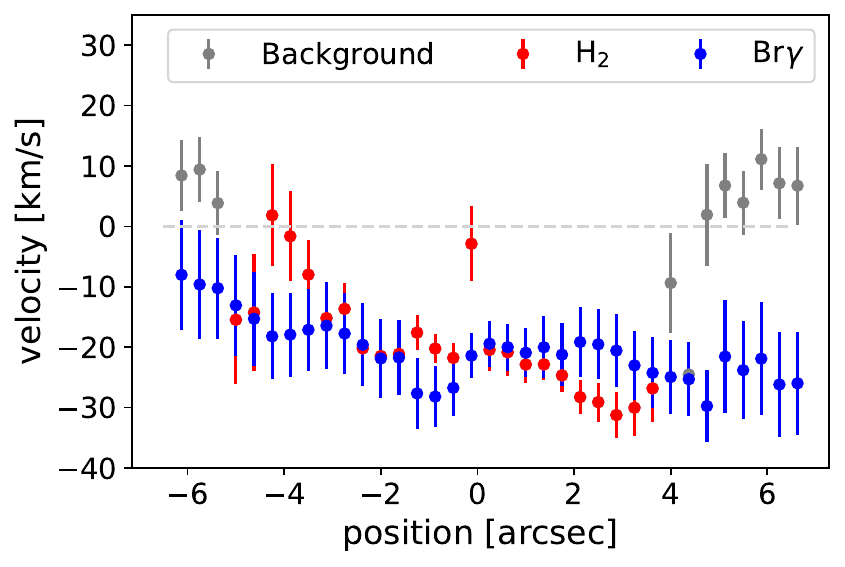}
    \caption{P-V slice that extends the entire length of the \brg\ outflow. }
    \label{fig:long_slice}
\end{figure}

\section{YSO spectra }\label{s:pcyc842}

\begin{figure*}
    \centering
    \includegraphics[width=\textwidth]{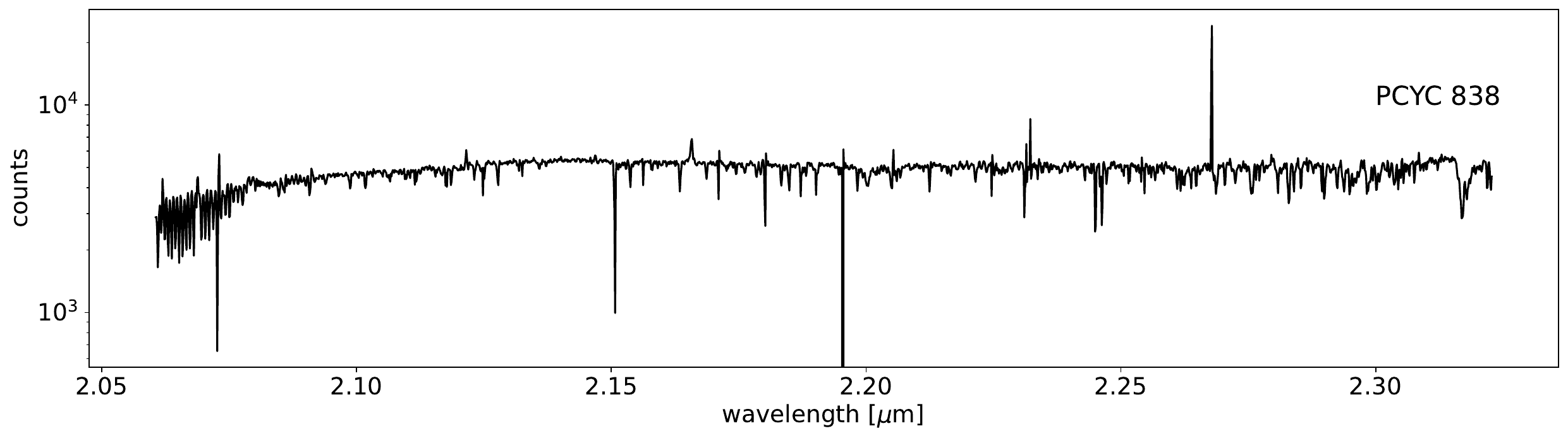}
    \caption{Spectrum of the PCYC~838~YSO located in the western limb of the HH~900 jet+outflow. Various emission lines are detected, including the Br$\gamma$ line, tracing ongoing accretion on the target.  }
    \label{fig:pcyc838_spec}
\end{figure*}

\begin{figure*}
    \centering
    \includegraphics[width=\textwidth]{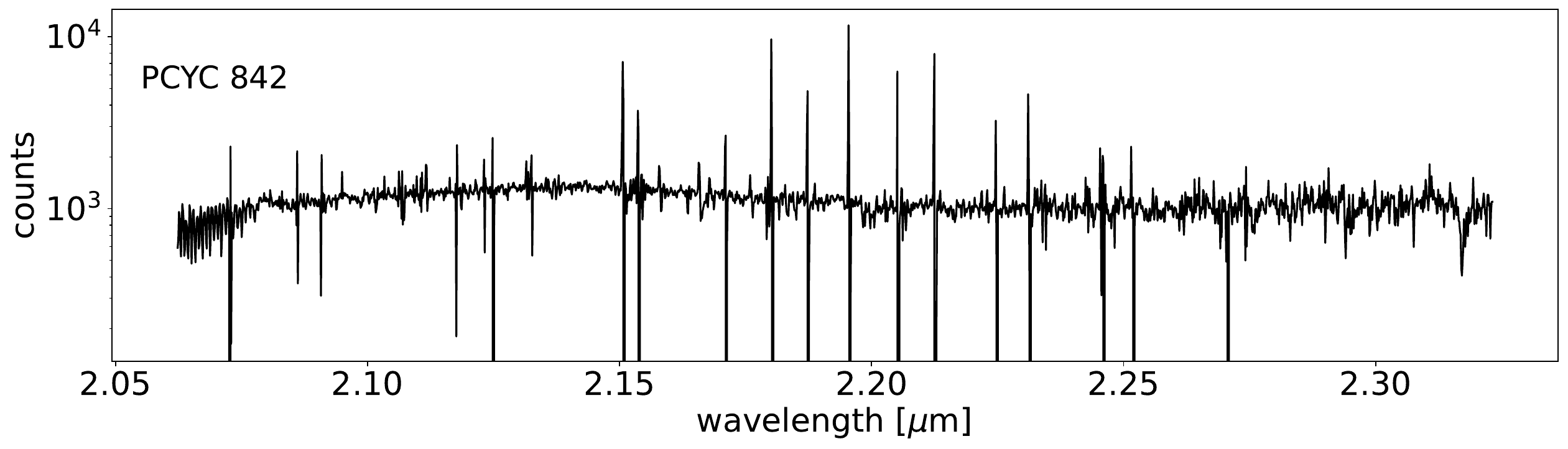}
    \caption{Spectrum of the star at the bottom of the globule \citep[PCYC~842 in][]{povich2011}. 
    }
    \label{fig:not_YSO_pcyc842}
\end{figure*}


\bsp	
\label{lastpage}
\end{document}